\documentclass[sigplan,twocolumn,nonacm]{acmart}

\renewcommand\footnotetextcopyrightpermission[1]{}
\usepackage{tikz}
\usepackage{amsmath}
\usepackage{amsfonts}

\usepackage{amssymb}
\usepackage[inline]{enumitem}
\usepackage{subcaption} 
\usepackage{booktabs}
\usepackage{textcomp}
\usepackage{hyperref}
\usepackage{algorithm}
\usepackage{natbib}
\usepackage{algpseudocode} 
\usepackage{graphicx} 
\usepackage{fontawesome5} 
\usepackage{threeparttablex} 
\usepackage{tabularx} 
\usepackage{tikz}
\usepackage[svgnames]{xcolor}
\usepackage{placeins} 
\usepackage[most]{tcolorbox}
\usepackage{soul}
\usepackage{float}
\allowdisplaybreaks

\definecolor{mypaperblue}{RGB}{9, 87, 162}

\newcommand*\numcircled[2][]{%
    \tikz[baseline=(char.base)]{
        \node[shape=circle, fill=black, text=white, inner sep=1pt, 
              minimum size=1.2em, font=\fontsize{9}{9}\selectfont #1] (char) {#2};
    }%
}

\newtheorem*{proof-no-qed}{Proof}

\AtBeginDocument{%
  }

\begin{document}

\title{Vulnerabilities in Partial TEE-Shielded LLM Inference with Precomputed Noise}
\author{Abhishek Saini}
\affiliation{%
  \institution{Rutgers, The State University of New Jersey}
  \city{Piscataway}
  \state{NJ}
  \country{USA}}
\email{abhishek.saini@rutgers.edu}

\author{Haolin Jiang}
\affiliation{%
  \institution{Rutgers, The State University of New Jersey}
  \city{Piscataway}
  \state{NJ}
  \country{USA}}
\email{hj462@scarletmail.rutgers.edu}

\author{Hang Liu}
\affiliation{%
  \institution{Rutgers, The State University of New Jersey}
  \city{Piscataway}
  \state{NJ}
  \country{USA}}
\email{hang.liu@rutgers.edu}

\renewcommand{\shortauthors}{Abhishek Saini, Haolin Jiang, and Hang Liu}

\begin{abstract}
The deployment of large language models (LLMs) on third-party devices requires new ways to protect model intellectual property. While Trusted Execution Environments (TEEs) offer a promising solution, their performance limits can lead to a critical compromise: using a precomputed, static secret basis to accelerate cryptographic operations. We demonstrate that this mainstream design pattern introduces a classic cryptographic flaw, the reuse of secret keying material, into the system's protocol. We prove its vulnerability with two distinct attacks: First, our attack on a model confidentiality system achieves a full confidentiality break by recovering its secret permutations and model weights. Second, our integrity attack completely bypasses the integrity checks of systems like Soter and TSQP. We demonstrate the practicality of our attacks against state-of-the-art LLMs, recovering a layer's secrets from a LLaMA-3 8B model in about 6 minutes and showing the attack scales to compromise 405B-parameter LLMs across a variety of configurations.
\end{abstract}

\ccsdesc[500]{Security and privacy~Key management}
\ccsdesc[500]{Security and privacy~Cryptanalysis and other attacks}
\ccsdesc[500]{Computing methodologies~Natural language processing}
\ccsdesc[500]{Security and privacy~Domain-specific security and privacy architectures}
\keywords{Confidential Computing, Machine Learning Security, Trusted Execution Environments (TEEs), Large Language Models (LLMs), Cryptographic Attacks, Key Reuse}


\maketitle

\section{Introduction}
\label{sec:intro}

The initial wave of Large Language Models (LLMs) has been dominated by centralized, API-based services \cite{openai_chatgpt_nodate, anthropic_meet_nodate, google_google_nodate}. However, this paradigm is rapidly giving way to a new frontier: local deployment of models outside the provider's direct control. This shift is not driven by a single factor, but by three distinct and essential requirements: First, enterprise users are increasingly running models on-premises to process sensitive, proprietary data, driven by concerns over exposing market secrets to competitors and the legal and compliance risks of sharing confidential information with third-party model providers \cite{openai_chatgptforenterprise_2025, sentinelone_chatgpt_2025}. Second, entire classes of mission-critical applications, such as autonomous driving and robotics, are non-negotiably edge-native; their need for real-time responsiveness and constant availability precludes any reliance on a remote server \cite{firoozi_foundation_2023, cui_survey_2024, tesla_autopilot_nodate, karumbunathan_nvidia_2022}. Finally, the proliferation of consumer devices like smartphones and smart assistants, now equipped with powerful AI accelerators, has created a demand for a new generation of responsive, always-on applications powered by local proprietary models \cite{apple_deploying_nodate, gupta_google_2023, wilhelm_amazon_2023}.

While each requirement is distinct, they converge on a single, critical security challenge: placing a multi-million dollar model, a core intellectual property asset, onto hardware not managed by the provider exposes it to unauthorized replication and reverse engineering \cite{sun_mind_2021}. This tension has led to a clear call from the research community for a ``middle path'' that delivers the full benefits of local execution without sacrificing the model provider's intellectual property~\cite{huang_position_2025, cheng_reclaiming_2025}.



One approach to securing inference relies on purely cryptographic methods like Homomorphic Encryption (HE), which allow computation on encrypted data. However, HE is often too slow for the demands of low-latency LLM inference~\cite{gilad-bachrach_cryptonets_2016, juvekar_gazelle_2018}. Consequently, hardware-based solutions using Trusted Execution Environments (TEEs) have emerged as a more practical alternative that offers a better balance of security and performance~\cite{hanzlik_mlcapsule_2019, ohrimenko_oblivious_2016}. However, the memory and performance constraints of TEEs make running an entire state-of-the-art LLM within a secure enclave impractical~\cite{lee_occlumency_2019, kim_vessels_2020}. While recent confidential computing architectures like Intel TDX, AMD SEV, Arm CCA, and NVIDIA's Confidential Computing are beginning to address these limitations, their practical adoption is hindered by limited availability or the persistent challenge of securing untrusted accelerators~\cite{kaplan_memory-encryption-white-paper_nodate, siby_guarantee_2024, abdollahi_early_2025, dong_evaluating_2025, zhu_confidential_2024, hunt_telekine_2020, vaswani_confidential_2023, volos_graviton_2018, chrapek_fortify_2024}. This has led to the mainstream paradigm of \textit{Partial TEE-Shielded Execution (PTSE)}, which offloads computationally intensive operations to powerful but untrusted hardware like Graphics Processing Units (GPUs), while the TEE retains a critical role in managing the overall secure execution of the model~\cite{tramer_slalom_2019, shen_soter_2022, li_translinkguard_2024, sun_shadownet_2023, stoica_berkeley_2017}. 



Early PTSE strategies focused on model partitioning, where the model is split, and only the most sensitive layers or components are executed inside the TEE \cite{mo_darknetz_2020, liu_mirrornet_2023, schlogl_ennclave_2020}. However, this strategy is flawed as it offers incomplete IP protection and underutilizes the accelerator \cite{li_teeslice_2024, sun_tsqp_2025}. 

\begin{table*}[h!]
\centering
\resizebox{\textwidth}{!}{%
\footnotesize
\begin{tabular}{@{}l p{2.47cm} p{4.5cm} p{4.5cm} p{5.5cm}@{}}
\toprule
\textbf{System} & \textbf{Security Goal} & \textbf{Masking Strategy (in TEE)} & \textbf{Offloaded Computation (on GPU)} & \textbf{Restoration Mechanism (in TEE)} \\ 
\midrule


ShadowNet \cite{sun_shadownet_2023} & 
Model Confidentiality &
Additive one-time pad, $M$, on the input feature map, $X$: $X' = X + M$. & 
Convolution with the transformed weights, $\hat{W}^T$, and the masked input, $X'$: $Y' = \text{Conv}(X', \hat{W}^T)$. & 
Restores $Y$ by subtracting the pre-computed noise effect and applying an inverse weight transformation: $Y = (Y' - \text{Conv}(M, W)) \cdot \Lambda^{-1}$. \\
\addlinespace

SLIP \cite{refael_slip_2024} & 
Model Confidentiality &
Additive one-time pad, $r_{i-1}$, on the input activation, $a_{i-1}$: $\tilde{a}_{i-1} = a_{i-1} + r_{i-1}$. & 
Matrix-vector multiplication with matrix $W_i^D$ and the masked activation, $\tilde{a}_{i-1}$: $\tilde{a}_i^D = W_i^D \cdot \tilde{a}_{i-1}$. & 
Restores $a_i^D$ by subtracting precomputed cancellation mask, $c_i$: $a_i^D = \tilde{a}_i^D - c_i = W_i^D a_{i-1}$, where $c_i = W_i^D r_{i-1}$. \\
\addlinespace

Soter \cite{shen_soter_2022} & 
Model Confidentiality &
Parameter Morphing: Multiplies the linear operator $F$ with a secret scalar blinding coin, $\mu$. & 
The morphed operator is executed on the GPU with plaintext input: $\mu \cdot F(X)$. & 
Restores $F(X)$ by multiplying the result by the reciprocal of the blinding coin: \newline $\mu^{-1} \cdot (\mu \cdot F(X)) = F(X)$. \\
\addlinespace

\begin{tabular}[t]{@{}l@{}}TransLinkGuard \\ (TLG) \cite{li_translinkguard_2024}\end{tabular} & 
Model Confidentiality &
Additive one-time pad, $m$, on the input activation $a'$, followed by a permutation $\rho_l$: $a'' = (a' + m)\rho_l$. & 
Matrix-vector multiplication with matrix, $W_3'$, and masked, permuted input, $a''$: $b'' = a''W_3'$. & 
Restores $b$ by subtracting the precomputed noise effect, $m W_b$: $(b'' - m W_3) = b$. \\
\addlinespace

\begin{tabular}[t]{@{}l@{}}Soter \cite{shen_soter_2022}, \\ TSQP \cite{sun_tsqp_2025}\end{tabular} & 
Computational \newline Integrity &
Challenge Injection: Injects a fingerprint vector $m'$, a random linear combination of a static basis of cornerstone inputs, $\{m_i\}$: $m' = \sum \alpha_i m_i$. &
The GPU computes the linear operation $F$ on a batch containing both genuine activations and the fingerprint challenge, producing $F(m')$. &
Result Verification: TEE computes the expected output, $F(m')$, using linearity of $F$ and precomputed cornerstone outputs, $\{F(m_i)\}$, $F(m') = \sum \alpha_i F(m_i)$ and compares it to GPU's result. \\
\addlinespace



\bottomrule
\end{tabular}%
} 
\caption{The mask-offload-restore pattern in practice across TEE-shielded inference systems.
\vspace{-.2in}
}
\label{tab:pattern-comparison-final}
\end{table*}

More recent PTSE approaches have evolved towards cryptographic obfuscation, which operates on a ``lock-and-key'' principle: The model is first obfuscated (locked) offline using a secret transformation, rendering it unusable by any party that does not possess the corresponding secret key. During online inference, the TEE's role is twofold: \ul{(i) Model confidentiality.} TEE guards the model's secrets by securely holding the key needed to de-obfuscate computations from the GPU. A common approach to obfuscation in this domain is permutation-based locking, which secretly shuffles a model's weights; the TEE then ensures the model only functions correctly by using the secret permutation key to properly order the data flowing into the locked model weights~\cite{li_translinkguard_2024, yuan_secure_2024, sun_shadownet_2023, xu_tempo_2024, zhang_groupcover_2024, wang_game_nodate}. Recent work has focused on scaling this approach to protect LLMs \cite{li_translinkguard_2024, refael_slip_2024, sun_tsqp_2025, wang_game_nodate}. \ul{(ii) Computational integrity.} The TEE serves to verify the integrity of computations offloaded to an untrusted hardware. This is a vital role, as an attacker with control over the host system could manipulate the GPU's computations. Such tampering can degrade the model's service quality, damaging the provider's reputation and eroding user trust. More maliciously, even an unprivileged, co-located attacker could aim for targeted manipulations \cite{kim_flipping_2014, hong_terminal_2019}, altering the computation in subtle ways to bypass security filters or cause specific misclassifications that benefit them \cite{rakin_t-bfa_2021, zhao_fault_2019}. To prevent this, systems like Soter~\cite{shen_soter_2022} and TSQP~\cite{sun_tsqp_2025} leverage the TEE hardware to inject hidden \textit{fingerprints}, known-answer challenges for which the TEE already knows the correct output, mixed in with real data. If the untrusted hardware returns a tampered result for these hidden checks, the integrity breach is detected, and the session can be determined as untrusted.


To achieve these distinct security goals, a common architectural pattern has emerged across a wide range of systems: a three-step \textit{Mask-Offload-Restore} protocol, which we detail and compare across several state-of-the-art systems in Table~\ref{tab:pattern-comparison-final}. The protocol proceeds as follows:

\vspace{-.1in}
\begin{enumerate}
    \item \textbf{Mask.} Inside the TEE, plaintext data is cryptographically masked. For confidentiality, this involves adding secret noise to an input; for integrity, it involves injecting a known-answer fingerprint challenge.
    \item \textbf{Offload.} The masked data is then offloaded to an untrusted hardware (e.g., GPU), which performs the computationally expensive linear operations.
    
    \item \textbf{Restore.} The result is returned to the TEE, which must efficiently restore the plaintext output by removing the noise's effect or verify the computation's integrity by checking the fingerprint's result.
\end{enumerate}

\begin{figure*}[t]
    \centering
    \includegraphics[width=\textwidth]{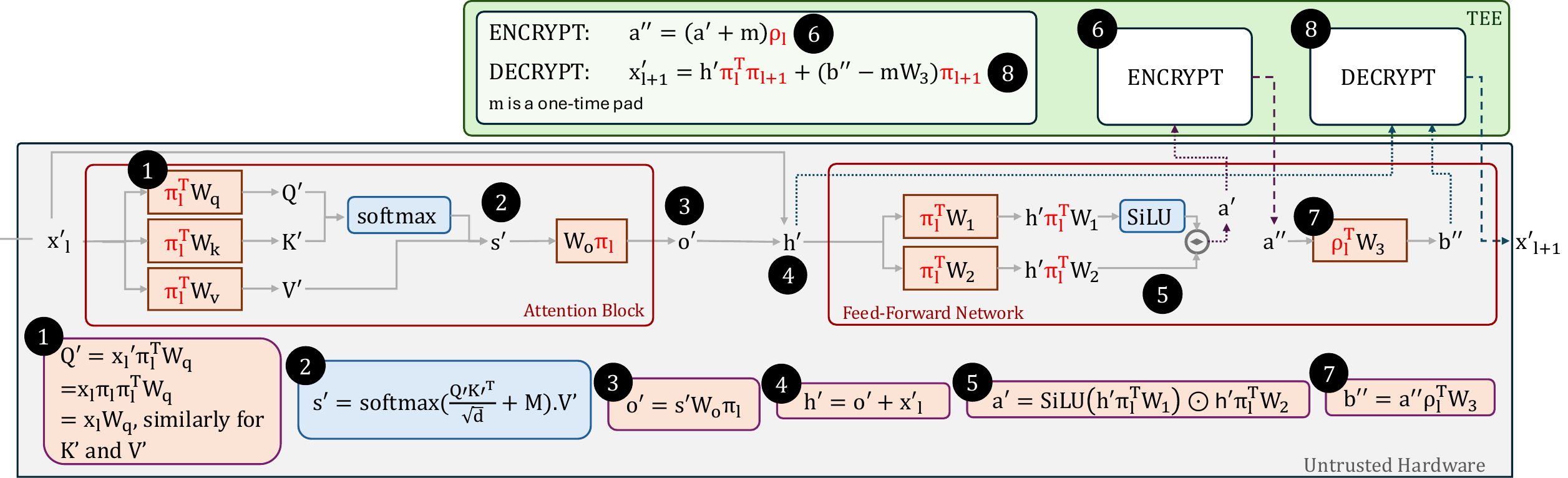}
    \caption{Activation flow in a Llama transformer layer secured using TLG~\cite{li_translinkguard_2024}. \vspace{-.1in}} 
\label{fig:architectural_overview}
\end{figure*}

The practical challenge arises because the TEE is responsible for more than just generating this randomness; it must also restore the original result by computing the precise effect of the noise. As we will demonstrate in Section \ref{sec:onthefly_vs_precomputation}, calculating this effect for truly random noise is computationally prohibitive. It would force the TEE to execute expensive \textit{On-the-Fly} operations, like full matrix-vector multiplications, creating an unacceptable performance bottleneck.

\vspace{.1in}
To circumvent this bottleneck, practical implementations are compelled to adopt a performance-driven shortcut: a \textit{precomputation}-based pattern. At startup, the TEE generates and stores a small, fixed set of $K$ secret noise vectors and their corresponding effects, which we call the precomputed static basis. During online inference, all ``random'' noise is then efficiently generated by combining these precomputed static basis vectors using freshly sampled random coefficients. For TEE-based model confidentiality, this efficiency-driven design is a necessity to scale to LLMs, while for others, like the integrity-checking mechanism in Soter, TSQP, it is adopted as a tempting shortcut for efficiency. 



\vspace{.1in}

In this work, we demonstrate that this efficiency-driven reliance on a precomputed static basis is a fundamental security flaw. The key insight is that the random noise generated this way is not truly random. Instead, it is permanently confined to a predictable, low-dimensional ($K$-dimensional) subspace spanned by the static basis vectors. This creates a critical vulnerability that allows an attacker to aggregate information over multiple interactions to algebraically break the system's security guarantees. This flaw is not an implementation bug or a side channel, but a systematic weakness inherent to the protocol's design.

\vspace{.1in}

We prove the severity of this flaw by developing attacks that compromise two distinct security guarantees: {model confidentiality system and computational integrity} framework. The success of our attacks on these two different real-world designs proves that the static basis is a versatile and fundamental vulnerability. Our primary contributions are:
\begin{itemize}
    \item The identification of the ``static secret basis'' pattern as a critical, performance-driven vulnerability in TEE-shielded inference systems.

    \item The development and formalization of a novel attack that breaks model confidentiality in systems like TLG~\cite{li_translinkguard_2024} by algebraically characterizing the noise subspace to recover secret permutations.

    \item The development of a more general attack that bypasses integrity protection in systems like Soter~\cite{shen_soter_2022}, TSQP~\cite{sun_tsqp_2025} by isolating the hidden noise subspace from other data through vector space intersection.
    


    \item An extensive evaluation that validates the effectiveness of our attacks. We demonstrate that the underlying vulnerability stems from a necessary performance trade-off. Our methods are robust and scalable to state-of-the-art LLM models across various configurations.
\end{itemize}

\textbf{Distinction from existing closely related attacks.}  While other protocol-level Partial TEE-shielded execution attacks exist, they typically rely on statistical inference and require a known public model to succeed \cite{wang_game_nodate, zhang_groupcover_2024, zhang_no_2023}. 
Our attack is fundamentally different from prior work. We exploit a mathematical flaw in the security protocol itself, assuming a perfect TEE. 
Our attack can perfectly isolate the reused secret basis, enabling attackers to fully obtain the protected LLM model and bypass integrity checks, demonstrating a new manifestation of key-reuse vulnerability in partial TEE-shielded executions. Please refer to Section~\ref{sec:related} for more discussions.

\section{Background and Threat Model}



\subsection{TEE-based Model Confidentiality}
\label{sec:bg:tee-based-auth}

Many TEE-based model confidentiality schemes operate on a lock-and-key authorization principle to enable secure computation on untrusted hardware \cite{bai_phantom_2025, xu_tempo_2024, shen_soter_2022, li_translinkguard_2024, sun_shadownet_2023, li_teeslice_2024, wang_game_nodate}. In this paradigm, the model's weights are first ``locked'' offline through an obfuscation technique, such as permutation, rendering them unusable without a corresponding secret key. This locked model is then deployed to the untrusted environment. The secret key, the permutation matrix in this case, is stored securely within the TEE. During inference, TEE uses the secret key to securely perform the secret transformation necessary for the correct functioning of the model.

As a concrete example of this mechanism, Figure \ref{fig:architectural_overview} illustrates a single layer of the Llama architecture secured using TLG~\cite{li_translinkguard_2024}. In the secured model, the weights are locked offline using layer-specific permutation matrices, $\pi_l$ and $\rho_l$. To an attacker, these locked weights appear random; they are unusable without the secret permutations, which are stored securely in the TEE. We denote secured activations with a prime (e.g., $x_l'$) to distinguish them from their plaintext counterparts ($x_l$).

The inference process works by applying corresponding permutations to the activations. An incoming activation $x'_l$, already permuted by the previous layer ($x'_l=x_l\pi_l$), is processed by the attention block. Due to the block's structure, the permutations on the weights and activations cancel out at step \numcircled{1}, ensuring the intermediate outputs are correct (e.g., $Q' = x'_l \pi_l^T W_q = (x_l \pi_l) \pi_l^T W_q = Q$).

The computed values $Q'$, $K'$, $V'$ are then combined to form the aggregated context, $s'$~(\numcircled{2}), which then gets multiplied by the output projection matrix, $W_o \pi_l$~\numcircled{3}. This gives the output of the attention block, $o'$. A residual connection is added to create $h'$ (\numcircled{4}), the input to the feed-forward network. At this stage, $h' = h \pi_l$ is permuted correctly for the first half of the feedforward network to compute $a'$ \numcircled{5}. Since $h'$ cancels out the permuted weights of the first half of the feed-forward network, $a'$ is the same as its plaintext version $a$. Therefore, $a'$ is not correctly permuted for the locked weight $W'_3 = \rho_l^T W_3$. To securely apply the required permutation $\rho_l$ without leaking it to an observer, the system uses a three-step ENCRYPT-LINEAR-DECRYPT sequence: the TEE first encrypts the activation $a'$ with a random one-time pad ($m$) before permuting it~(\numcircled{6}), the GPU performs the linear computation~\numcircled{7}, and TEE decrypts the result by removing the noise effect ($mW_3$) to recover the correct output~(\numcircled{8}).

To ensure the cryptographic security of one-time pad masking, all computations are performed over a finite field. Following standard practice, all floating-point activations and weights are first quantized to fixed-point integers. These integers are then treated as elements of the finite field $\mathbb{F}_P$, where $P$ is a large prime chosen to prevent wrap-around errors. This step is what allows the random mask m to function as a secure one-time pad, providing a formal basis for the system's security guarantees \cite{tramer_slalom_2019}.

\subsection{TEE-based Computational Integrity}
\label{sec:soter-mechanism}

\begin{figure}[t]
  \centering
  \includegraphics[width=0.95\linewidth]{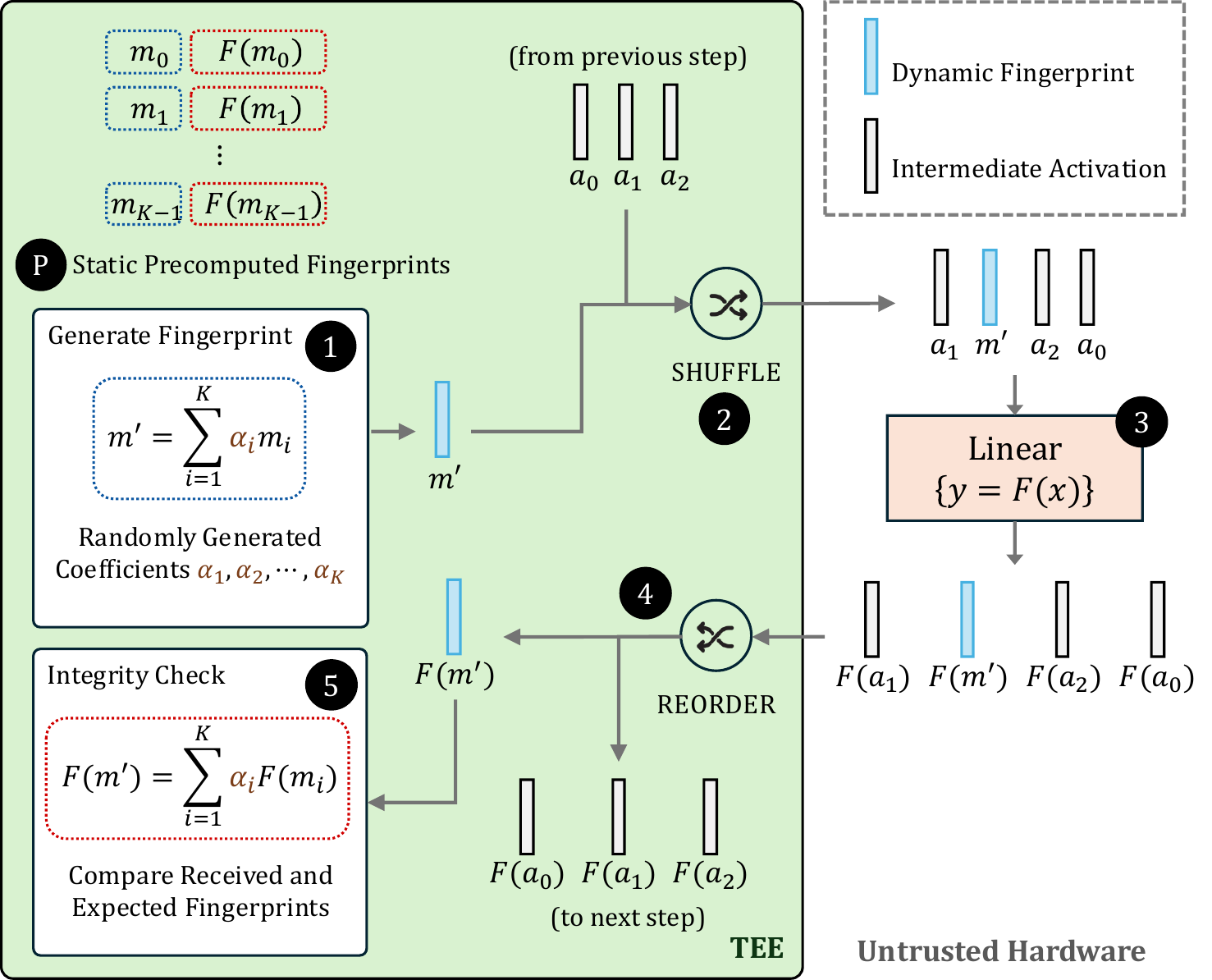}
  \caption{Soter's mechanism for generating integrity fingerprints from a static basis of precomputed cornerstones.\vspace{-.2in}}
  \label{fig:soter-mechanism}
\end{figure}

Model's computational integrity is sometimes protected using a technique called \emph{oblivious fingerprinting}~\cite{shen_soter_2022, sun_tsqp_2025}. The core idea is to mix known-answer fingerprints, challenges for which the TEE already knows the correct output, with the genuine user inputs sent to the GPU. To prevent an attacker from simply detecting and ignoring these checks, the fingerprints are designed to be statistically indistinguishable from normal inputs. By verifying the GPU's output for these hidden fingerprints, the TEE can detect any tampering. 

Figure~\ref{fig:soter-mechanism} exemplifies this mechanism used in Soter~\cite{shen_soter_2022} and TSQP~\cite{sun_tsqp_2025}. 
Beginning with a one-time offline setup where the TEE creates and stores a static basis of $K=10$ cornerstone fingerprint pairs, $\{m_i, F(m_i)\}$ (\numcircled{P}). During each online inference, the TEE generates a dynamic challenge, $m'$, as a random linear combination of the cornerstone inputs ($m' = \sum_{i=1}^{K} \alpha_i m_i$) (\numcircled{1}). This challenge is then shuffled with genuine activations and sent to the GPU (\numcircled{2}), which applies the operator $F$ to the entire batch \numcircled{3}. Upon receiving the results, the TEE unshuffles them (\numcircled{4}) and verifies the GPU's work by comparing the returned fingerprint output, $F(m')$, against the expected result it computes locally: $F(m')_{expected} = \sum_{i=1}^{K} \alpha_i F(m_i)$ (\numcircled{5}). A mismatch signifies an integrity breach and aborts the process.

\subsection{Threat Model}
\label{sec:threat_model}
We consider a setting where a defender (model provider) deploys a proprietary model on a user's device, which is controlled by a powerful adversary. The defender's goal is to guarantee two security properties for the deployed model: model confidentiality, to protect the intellectual property of its weights and parameters, and computational integrity, to verify the correctness of any computation offloaded to untrusted hardware.

We assume the TEE itself is a secure black box, free from physical or hardware side-channel attacks. All components outside the TEE including the host operating system, system memory, and the GPU accelerator are considered untrusted and are fully controlled by the adversary. The adversary's capabilities are therefore formidable: they can observe, intercept, and modify all communication between the TEE and the untrusted GPU. They know the complete architecture of the deployed model and the full details of the security protocol being used, but they do not have access to the secret keying material (e.g., secret permutations, static noise basis) held within the TEE. The adversary's objective is to break the defender's security guarantees. Aligned with our two primary case studies, we define two concrete goals for the adversary: \textbf{(1)} to compromise confidentiality by recovering the secret keys needed to fully reconstruct the original model (as in a TLG-like system), or \textbf{(2)} to compromise integrity by identifying and bypassing the verification mechanism to tamper with results without detection (as in Soter).



\section{Key Observations on the Trade-off between Security vs Computation Cost}
\label{sec:onthefly_vs_precomputation}


\subsection{Key Observations}

\begin{tcolorbox}[colback=black!5, colframe=black!75, boxrule=1pt, left=3pt, right=3pt,top=0pt,bottom=0pt]
{\small
\textbf{Observation 1:} TLG authorization process critically hinges on the TEE's ability to efficiently calculate the noise effect $mW_3$.
}
\end{tcolorbox}

Analyzing the workflow of TLG, we observe that the computations inside the TEE (\numcircled{6} and \numcircled{8} in Figure~\ref{fig:architectural_overview}) can lead to a bottleneck. In particular, the calculation of the noise effect, $mW_3$, should be efficient since it has the largest computational footprint. In the meantime, as the model continues to grow (e.g., LLM), and the deployment of these LLMs on mobile devices becomes prevalent~\cite{sorensen_leftoverlocals_2024, li_translinkguard_2024}, the memory cost mainly surrounding $W_3$ could also be a concern (Table~\ref{tab:model_architectures}). 

\begin{tcolorbox}[colback=black!5, colframe=black!75, boxrule=1pt, left=3pt, right=3pt,top=2pt,bottom=2pt]
{\small
\textbf{Observation 2:} Soter's design relies on the use of a static basis of precomputed fingerprints to efficiently calculate the fingerprint's expected output $F(m')$.
}
\end{tcolorbox}

This observation highlights a fundamental tension in Soter's design. For the integrity mechanism to be secure, the dynamically generated fingerprints must be statistically indistinguishable from genuine activations. However, for the mechanism to be efficient, Soter relies on this static, precomputed basis to enable rapid integrity checks, a point acknowledged in the original paper~\cite{shen_soter_2022}. As we show later in the paper, this reliance on a static basis forces all generated fingerprints into a predictable, low-dimensional subspace, making them computationally distinguishable from genuine inputs and undermining the system's security guarantee. 


In summary, the mechanisms in our two case studies, TLG and Soter, share a critical dependency: the need for an efficient TEE-based calculation of a noise effect ($mW_3$, $F(m')$). 
As we will detail in Section~\ref{subsec:detailed}, there are two main strategies to compute this effect: an on-the-fly method that generates fresh randomness for each query, versus a more performant precomputation-based approach that relies on a static, pre-loaded table of noise vectors.
However, the prohibitive memory and latency costs of the on-the-fly method force designers to adopt the precomputation-based approach, as shown in our subsequent detailed analysis of a TLG-like system.

\subsection{Detailed Analysis}
\label{subsec:detailed}

This section uses TLG to explain the on-the-fly vs precomputed approaches with the memory and latency constraints in consideration.

\begin{figure}[t]
  \centering
  \includegraphics[width=0.95\linewidth]{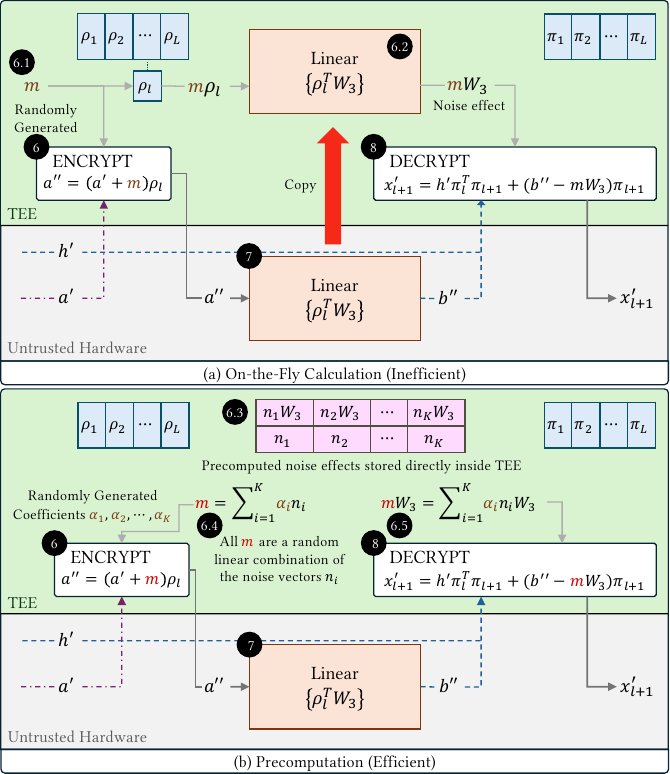}
  \caption{A comparison of two methods for calculating noise effects inside the TEE. (a) The on-the-fly approach.
  (b) The precomputation approach.
  }
  \label{fig:onthefly_vs_precomputation}
\end{figure}

\textbf{On-the-Fly Approach} employs an ENCRYPT-LINEAR-DECRYPT sequence to authorize computations on the untrusted platform. As illustrated in Figure~\ref{fig:onthefly_vs_precomputation}(a), the TEE first ENCRYPTs (\numcircled{6}) an activation by masking it with a random One-Time Pad (OTP), $m$. After the LINEAR computation is performed on the GPU (\numcircled{7}), the result re-enters the TEE. The DECRYPT function (\numcircled{8}) then recovers the correct output by subtracting the effect of the initial noise, $mW_3$. The method used to obtain this noise effect, $mW_3$, \numcircled{6.2} is the most time-consuming operation across the entire sequence.

The on-the-fly approach first randomly generates the noise vector $m$ (\numcircled{6.1}). To calculate the corresponding noise effect $mW_3$, the entire permuted weight matrix, $\rho_l^T W_3$, must be copied from untrusted memory into the TEE (\numcircled{2'}), where the matrix-vector multiplication is performed (\numcircled{6.2}).
However, this approach is inefficient as $W_3$ imposes (i) high memory demand and (ii) significant processing latency in the TEE.

\begin{table}[h]
\centering
\tiny
\begin{tabularx}{\columnwidth}{@{}l >{\hsize=0.6\hsize}X >{\hsize=1.4\hsize}X X@{}}
\toprule
\textbf{TEE Technology} & \textbf{Architectural Model} & \textbf{Available Memory} & \textbf{Bottleneck} \\
\midrule
\textbf{SGX v1} & Application \newline Isolation & Small, static EPC (e.g., 128 MB) & Memory and compute. \\
\addlinespace
\textbf{SGX v2} & Application \newline Isolation & Small (e.g., 128 MB) on clients; Large (GBs) on servers. & Memory and compute on clients; Compute on servers. \\
\addlinespace
\textbf{ARM TrustZone} & System \newline Partitioning & Small, fixed DRAM (16-64 MB) & Memory and compute. \\
\addlinespace
\textbf{AMD SEV-SNP} & Hardware- \newline Isolated VM & Large, dynamic protected memory (GBs) & Primarily compute. \\
\addlinespace
\textbf{Intel TDX} & Hardware- \newline Isolated VM & Large, dynamic protected memory (GBs) & Primarily compute. \\
\addlinespace
\textbf{ARM CCA} & Hardware- \newline Isolated VM & Large, dynamic protected memory (GBs) & Primarily compute. \\
\bottomrule
\end{tabularx}
\caption{Memory constraints comparison across representative TEE technologies. \vspace{-.2in}
}
\label{tab:tee_comparison}
\end{table}

\begin{table}[h]
\centering
\resizebox{\columnwidth}{!}{%
\begin{tabular}{@{}lrrrr@{}}
\toprule
\textbf{Model Family} & \textbf{Model Size} & \textbf{FFN Dim ($d_{\text{in}}$)} & \textbf{Hidden Dim ($d_{\text{out}}$)} & \textbf{$W_3$ Memory} \\ \midrule
LLaMA-3 & 8B & 14,336 & 4,096 & $\approx 224$~MB \\
Gemma 3 & 27B & 21,504 & 5,376 & $\approx 441$~MB \\
LLaMA-3 & 70B & 28,672 & 8,192 & $\approx 896$~MB \\
Mistral Large 2 & 123B & 28,672 & 12,288 & $\approx 1.31$~GB \\
LLaMA-3.1 & 405B & 53,248 & 16,384 & $\approx 3.25$~GB \\ \bottomrule
\end{tabular}%
} 
\caption{Architectural parameters and memory footprint of the $W_3$ matrix in recent open-source LLM families. \vspace{-.2in}
}
\label{tab:model_architectures}
\end{table}

\textit{(i) Memory demand} per layer—given by $d_{\text{in}} \times d_{\text{out}} \times s_{\text{dtype}}$ (matrix dimensions times datatype size)—often exceeds the capacity of memory-constrained TEEs. In LLaMA-3-8B model, for example, $W_3$ requires 112 MB in bfloat16, which doubles to 224 MB when converted to 32-bit integers required for \numcircled{6.2}, getting close to the memory limits of typical SGX and ARM TrustZone platforms (see Table~\ref{tab:tee_comparison}). Moreover, this footprint grows substantially for larger models (Table~\ref{tab:model_architectures}), making memory a critical bottleneck for on-the-fly approaches. We also want to mention that in modern server-based TEEs (e.g., Intel SGX v2, Intel TDX, AMD SEV-SNP, ARM CCA), one can have 10s of GB of memory that can hold large LLMs.

\textit{(ii) Significant latency} is the key concern that derails the on-the-fly approach. Particularly, the latency $T_{\text{fly}}$ contains two parts: (i) memory copy latency ($\frac{d_{\text{in}} \times d_{\text{out}} \times s_{\text{dtype}}}{BW_{\text{tee}}}$) and (ii) CPU computation latency ($\frac{2 \times d_{\text{in}} \times d_{\text{out}}}{\text{FLOPS}{\text{cpu}}}$), where $BW_{\text{tee}}$ is the TEE memory bandwidth and $\text{FLOPS}_{\text{cpu}}$ the in-enclave CPU performance. 
Therefore, $T_{\text{fly}} = \left( \frac{s_{\text{dtype}}}{BW_{\text{tee}}} + \frac{2}{\text{FLOPS}_{\text{cpu}}} \right) \times d_{\text{in}} \times d_{\text{out}}$.
%
Our benchmarks show that for the LLaMA-3-8B model, CPU computation alone exceeds $70$ ms per layer, resulting in more than $2$ s per token in a 32-layer model, which is too slow for low-latency applications. As $W_3$ increases in larger architectures (Table~\ref{tab:model_architectures}), the latency scales proportionally and becomes even more severe. For the LLaMA-3.1-405B model, the CPU computational latency approaches 1 second per layer. This will render a per-token latency up to 32 seconds. 


\textbf{Precomputation-based Approach}, shown in Figure~\ref{fig:onthefly_vs_precomputation}(b), addresses these bottlenecks by replacing costly matrix multiplications with lightweight linear combinations.
Instead of loading $W_3$ into the TEE, a small table of $K$ basis noise vectors $\{n_1,\dots,n_K\}$ and their precomputed effects $\{n_1W_3,\dots,n_KW_3\}$ (\numcircled{6.4}) is prepared at startup.
During inference, the TEE samples random coefficients $\{\alpha_1, ..., \alpha_K\}$ to construct both the noise vector $m=\sum_{i=1}^K \alpha_i n_i$ (\numcircled{6.4}) and its effect $mW_3 = \sum_{i=1}^K \alpha_i \times (n_i W_3)$ (\numcircled{6.5}), thereby avoiding on-the-fly multiplications.

This design reduces both memory and latency to practical levels. The memory footprint is now determined by the size of this precomputed table, which must store the $K$ basis noise vectors (each of dimension $d_{in}$) and their $K$ resulting effects (each of dimension $d_{out}$). The memory cost becomes $K\times(d_{\text{in}}+d_{\text{out}})\times s_{\text{dtype}}$, which is orders of magnitude smaller since $K \ll d_{\text{out}}$. 
Similarly, the latency is also significantly lower than in the on-the-fly approach.
For the LLaMA-3-8B model with $K=10$, the per-layer footprint drops from $\sim$224 MB to $\sim$0.7 MB (>300$\times$ reduction), while latency decreases from 70 ms to 0.23 ms per layer ($\sim$300$\times$ reduction). 

\textbf{The choice of $K$.} 
Intuitively, increasing $K$ enhances security by introducing more randomness. In practice, however, $K$ is limited by both memory and latency constraints:

\noindent First, the precomputed table across all $L$ layers must fit within the TEE's memory budget $M_{\text{tee}}$, giving: 
$$ L \times K_0 \times (d_{\text{in}} + d_{\text{out}}) \times s_{\text{dtype}} \le M_{\text{tee}}, K_0 \le \frac{M_{\text{tee}}}{L \times (d_{\text{in}} + d_{\text{out}}) \times s_{\text{dtype}}}.$$ 
Second, the in-enclave computation across all $L$ layers must not exceed the overall latency budget ($T_{\text{budget}}$), giving:
$$ L \times \frac{2 \times K_1 \times (d_{\text{in}} + d_{\text{out}})}{\text{FLOPS}_{\text{cpu}}} \le T_{\text{budget}}, K_1 \le \frac{T_{\text{budget}} \times \text{FLOPS}_{\text{cpu}}}{2 \times (d_{\text{in}} + d_{\text{out}})}.$$

In a practical deployment, $K$ must satisfy both constraints, making the true maximum basis size $K_{\max} = \min(K_0, K_1)$. 
For LLaMA-3 8B on a 128 MB TEE, the memory constraint alone bounds $K_{\max} \approx 56$ in a TLG-like system, with the actual value $K$ necessarily set even lower to reserve space for runtime data and secrets.

The memory and latency constraints force $K$ to be small, which creates a low-rank vulnerability that our attacks exploit.
These constraints are inherent to TEEs and therefore apply to all precomputation-based approaches.
For example, Soter uses a static basis of $K=10$ fingerprints to achieve integrity checking with low latency and acceptable memory overhead.
To demonstrate the severity and generality of this flaw, we next present two attacks: first on a TEE-based model confidentiality protocol (Section \ref{sec:direct_subspace_attack}), and second on a TEE-based integrity checking mechanism (Section \ref{sec:cross_attack_soter}).

\section{Attack on TEE-based Model Confidentiality}
\label{sec:direct_subspace_attack}
This section presents an attack on the TEE-based model confidentiality protocol of a TLG-like system that uses a static precomputed basis of noise vectors due to the practical constraints discussed in Section~\ref{sec:onthefly_vs_precomputation}. The attack ultimately results in the full leak of the model weights.

\subsection{Attack Formalization}

\begin{figure}[t]
  \centering
  \includegraphics[width=\linewidth]{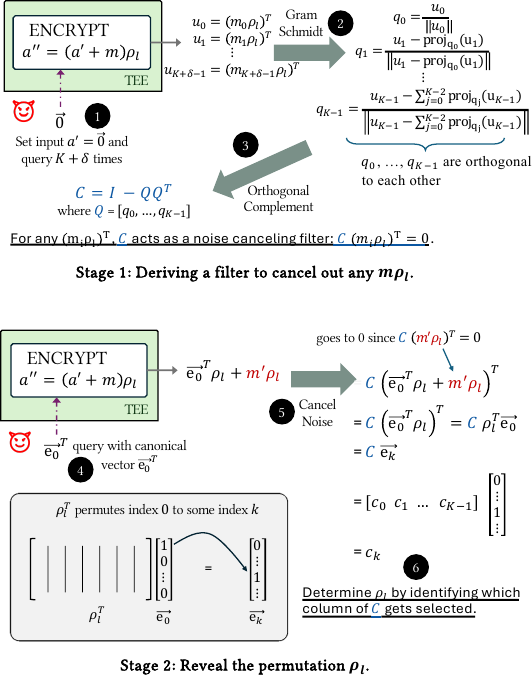}
  \caption{An illustration of the two-stage attack to reveal $\rho_l$ of \numcircled{6} in Figure~\ref{fig:onthefly_vs_precomputation}(b). 
  \vspace{-.18in}
  }
  \label{fig:direct_attack}
\end{figure}


The attack compromises the model layer by layer. For a given layer $l$, a full break requires recovering its two secret permutation matrices: $\rho_l$ (\numcircled{6} in Figure~\ref{fig:onthefly_vs_precomputation}) and $\pi_{l+1}$ (\numcircled{8} in Figure~\ref{fig:onthefly_vs_precomputation}), which authorizes the output for the subsequent layer. 
Note that as part of this layer-wise procedure, permutation $\pi_l$ is not a target when attacking layer $l$, as it would have already been compromised during the attack on the preceding layer, $l-1$. Additionally, for the initial layer ($l=1$), $\pi_1$ is recovered directly from the model's input stage. An attacker can observe the permuted token embeddings, which must be transformed by $\pi_1$ to align correctly with the first layer's locked weights. With $\pi_1$ known, the attack can proceed. The process for recovering each layer's secret matrices ($\rho_l$ and $\pi_{l+1}$) then consists of two main steps:
\begin{itemize}
    \item \textbf{Stage 1: deriving a filter to cancel out $m\rho_l$.} The attacker first learns the basis of the $K$-dimensional permuted noise subspace that spans $m\rho_l$. This is achieved by querying the ENCRYPT function $K + \delta$ times with a zero-vector input to gather enough samples of the noise (Stage 1 of Figure~\ref{fig:direct_attack}).
    \item \textbf{Stage 2: Reveal the secret permutation $\rho_l$.} With the noise-canceling filter from Stage 1, $m\rho_l$ can be canceled out, leaving only $a'\rho_l$. The attacker queries the ENCRYPT function $d$ more times, once for each canonical basis vector (e.g., $\vec{e}_1, ~\vec{e}_2, \cdots, ~\vec{e}_d$) to reveal where each input dimension is permuted, thereby reconstructing the entire permutation matrix $\rho_l$ (Stage 2 of Figure~\ref{fig:direct_attack}).
\end{itemize}

\textbf{Reveal $W_3$,  $\pi_{l+1}$, $W_q$, $W_k$, $W_v$, $W_o$, $W_1$ and $W_2$.}
After $\rho_l$ is revealed, we can reveal $W_3$ in \numcircled{7}. 
Finally, one can set h'=0 to reveal $\pi_{l+1}$ in \numcircled{8} in Figure~\ref{fig:onthefly_vs_precomputation}(b) using the same procedure as that used for revealing $\rho_l$. This reveals $W_q$, $W_k$, $W_v$, $W_o$, $W_1$ and $W_2$ for layer $l+1$. For a model with $L$ transformer layers, an attacker can repeat this step for all layers to recover all secret permutations and fully unlock the model.



\subsubsection{Stage 1: Extracting the Subspace Basis for $m\rho_l$}

The first stage of the attack is to characterize the $K$-dimensional subspace containing the permuted noise vectors $m \rho_l$ and construct a noise-canceling filter. This is possible because the noise vector $m$ is not truly random but is always a linear combination of $K$ secret basis vectors, $m = \sum_{i=1}^K \alpha_i n_i$. Consequently, the output $m \rho_l$ is always confined to a predictable, low-dimensional subspace. The attacker proceeds with the following steps (see Figure~\ref{fig:direct_attack}):

\begin{enumerate}
    \item \numcircled{1} Query for noise samples: The attacker queries the ENCRYPT function $K+\delta$ times using a zero-vector input, $a' = \vec{0}^T$. The function returns $K+\delta$ permuted noise samples as $1 \times d$ row vectors, $m_i\rho_l$. The attacker transposes these into a set of $d \times 1$ column vectors, which we denote as $\{ u_i = (m_i\rho_l)^T \}_{i=0}^{K+\delta-1}$.

    \item \numcircled{2} Derive orthonormal basis: Using the collected column vectors $\{u_i\}$, the attacker applies the Gram-Schmidt process to compute an orthonormal basis for the subspace spanned by permuted noise, $m \rho_l$. This results in a $d \times K$ matrix $Q = [q_0, q_1, \dots, q_{K-1}]$, whose columns span the subspace.

    \item \numcircled{3} Construct noise-canceling filter: The attacker uses this basis to construct a $d \times d$ projection matrix, $C = I - QQ^T$. This matrix projects any vector onto the orthogonal complement of the noise subspace, acting as a perfect noise-canceling filter. For any collected noise vector $u_j = (m_j\rho_l)^T$, it holds that $Cu_j = 0$.
\end{enumerate}


\subsubsection{Stage 2: Extracting the Permutation Matrix $\rho_l$}

In the second stage, the attacker uses the noise-canceling filter $C$ to reveal the secret permutation $\rho_l$ one mapping at a time. The reveal of the first mapping using the first canonical vector $\vec{e}_0 = [1, 0, 0, ...]^T$ is illustrated in Figure~\ref{fig:direct_attack}: 

\begin{enumerate}
    \item \numcircled{4} Query with a canonical vector: The attacker queries the ENCRYPT function with the first canonical basis vector as input, $a' = \vec{e}_0^T$. The function returns the row vector $(\vec{e}_0^T\rho_l + m'\rho_l)$.

    \item \numcircled{5} Apply the filter: The attacker transposes this output into a column vector, $v = (\vec{e}_0^T\rho_l)^T + (m'\rho_l)^T$, and applies the projection matrix:
    \begin{align*}
        Cv &= C((\vec{e}_0^T\rho_l)^T + (m'\rho_l)^T) \\
           &= C(\vec{e}_0^T\rho_l)^T + C(m'\rho_l)^T.
    \end{align*}
    Since the noise term $C(m'\rho_l)^T$ is nullified by the filter, the result is only the projected signal:
    \[
        Cv = C(\vec{e}_0^T\rho_l)^T.
    \]

    \item \numcircled{6} Reveal the mapping: The term $(\vec{e}_0^T\rho_l)^T = \rho_l^T \vec{e}_0$ is itself a canonical column vector, which we can call $\vec{e}_k$. The final result is thus $C\vec{e}_k$, which is equivalent to the $k$-th column of the matrix $C$. By observing the resulting vector and matching it to the corresponding column of the matrix $C$, the attacker can determine the index $k$. This reveals one mapping of the secret permutation: $\rho_l^T$ maps index $0$ to index $k$.
\end{enumerate}

By repeating this process for all other canonical vectors ($\vec{e}_1, \dots, \vec{e}_{d-1}$), the attacker can recover the entire secret permutation matrix $\rho_l$. This two-stage process of characterizing the noise subspace and then filtering it to isolate the signal is formalized in Appendix \ref{app:algorithms} - Algorithm \ref{alg:perm_recovery_tlg_for_paper}. 



\subsection{Discussion on sampling strategy, $K$ and $\delta$}
Even if the TEE uses a complex strategy to combine the noise vectors, such as combining a random subset of the basis vectors for each query, the output $m \rho_l$ is still confined to the same $K$-dimensional subspace, making the system vulnerable. We experimentally demonstrate this in Section~\ref{sec:eval:robustness}. 

$K$ is a protocol parameter chosen by the model provider during the offline noise table precomputation step.
Our threat model assumes the attacker has full knowledge of the protection protocol, including $K$, except for the secret keys and permutations. However, even if the attacker does not know $K$, it can be found empirically by collecting outputs and observing the rank of the resulting vector set until it stabilizes. 
We confirm the practicality of the attack under unknown $K$ (Section \ref{sec:eval:k_discovery}), demonstrating that the attack succeeds even when $K$ is initially unknown. Discovery of $K$ is possible because the rank is guaranteed to stabilize at $K$ since $m \rho$ is generated from a fixed $K$-dimensional subspace. 

Our protocol assumes that a linear combination of all $K$ vectors is used to generate the noise mask. Therefore, by querying $K+\delta$ times, the attacker observes enough samples of the permuted noise subspace to derive its basis, $Q$ (\numcircled{2} in Figure~\ref{fig:direct_attack}). The small constant $\delta > 0$ is chosen to avoid the pathological case when some of the random combinations are linearly dependent by pure chance. 

For a $K \times K$ matrix whose entries are chosen uniformly at random from a finite field with $P$ elements, the probability of it being rank deficient (linearly dependent) is:
\[
P(\text{rank deficient}) = 1 - \prod_{i=1}^{K} \left( 1 - \frac{1}{P^{i}} \right).
\]
A derivation for this result has been provided in Appendix \ref{app:derivation_rank_def}.
In our implementation of LLaMA-3-8B model, we used $P = 2^{31}-1$ for which this probability is negligible. So $\delta$ can be something small like 1 or 2. 

\section{Attack on TEE-based Computational Integrity}
\label{sec:cross_attack_soter}

In this section, we show an attack that fully compromises a TEE-based integrity checking mechanism. In particular, we use Soter~\cite{shen_soter_2022} as an example to explain the attack, although the same mechanism is also used in TSQP~\cite{sun_tsqp_2025}.

\begin{figure}[t]
  \centering
  \includegraphics[width=\linewidth]{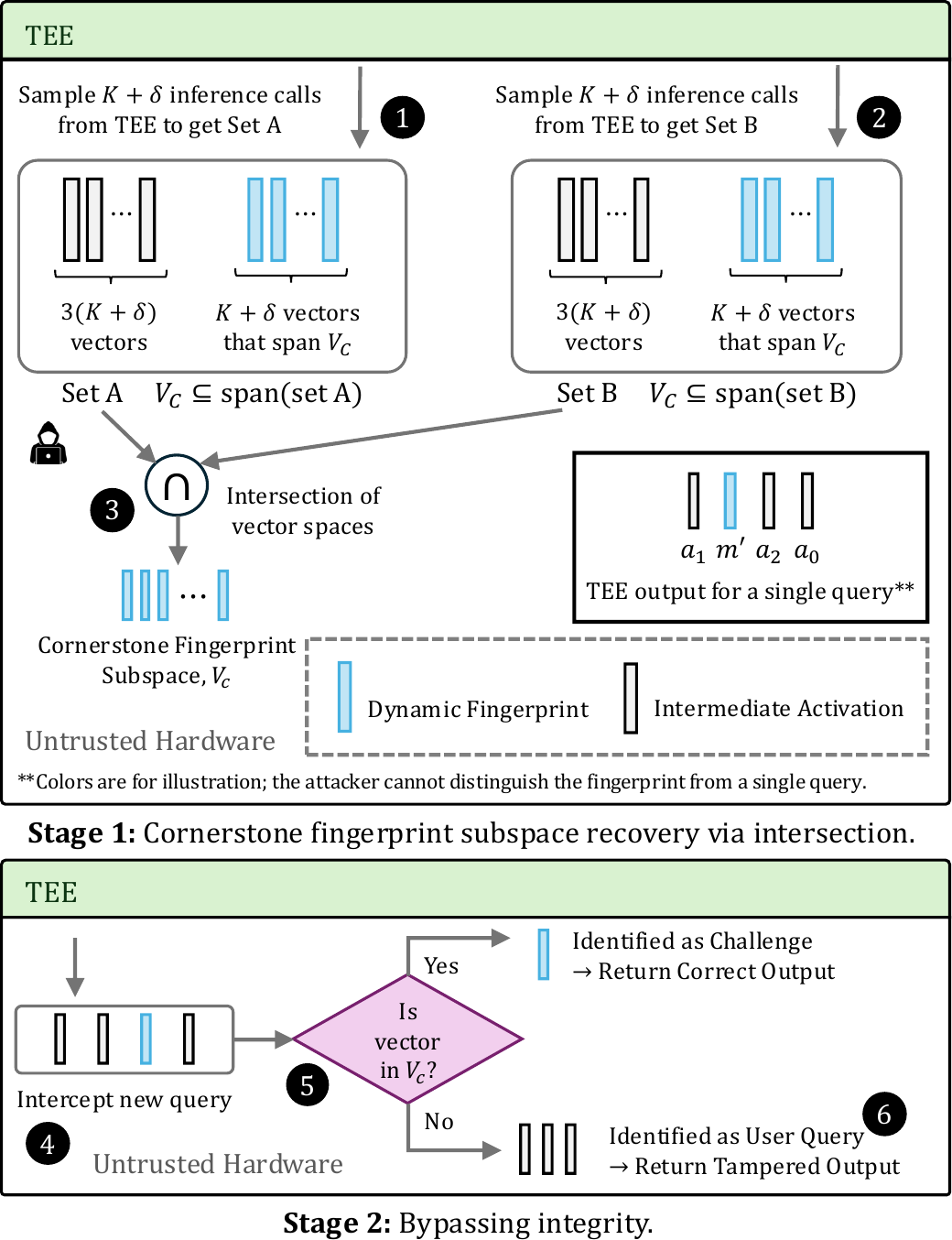}
  \caption{An illustration of the two-stage computation integrity attack on Soter. \textbf{Stage 1: Subspace Recovery.} The attacker collects two independent sets of vectors from the TEE's output, Set A and Set B. By computing the intersection of the vector spaces they span, the attacker recovers the secret cornerstone fingerprint subspace ($V_C$). \textbf{Stage 2: Integrity Bypass.} The attacker intercepts a new query (4), uses the recovered subspace as a filter to identify the challenge fingerprints.}
  \label{fig:intersection_attack}
\end{figure}

\subsection{Attack Formalization}
\label{sec:soter-attack}

Soter protects computational integrity by mixing hidden, known-answer challenge fingerprints among genuine activations. By verifying the GPU's computation on these challenges, the TEE can detect any tampering. The attacker's goal, therefore, is to distinguish these challenge fingerprints from the genuine data. Doing so would allow the attacker to pass the integrity check by computing the challenges correctly while selectively tampering with genuine activations. The attack works in two stages as illustrated in Figure~\ref{fig:intersection_attack}:

\paragraph{Stage 1: Revealing the subspace for $m_i$.}
In the first stage, the adversary learns a basis for the $K$-dimensional cornerstone fingerprint subspace, which we denote as $V_C$.

\paragraph{Stage 2: Identifying the fingerprint to bypass the integrity check.}
Armed with a basis for $V_C$, the adversary can now distinguish the fingerprints from genuine activations. Using this knowledge, the attacker only tampers the genuine activations while correctly computing the integrity checks. 

\subsubsection{Stage 1: Revealing the subspace for $m_i$}

The attacker begins by passively monitoring the TEE-GPU interface to collect two independent sets of outputs, Set A (\numcircled{1}) and Set B (\numcircled{2}). Each set itself is generated from $K+\delta$ independent inference calls. For Soter, $K=10$. Because each challenge is generated from a fresh, random linear combination of all cornerstone fingerprints, $K+\delta$ calls for each set ensure that the cornerstone fingerprint subspace is a subspace of the vector space spanned by each set. The inference calls are kept independent of each other to avoid any linear dependence within the genuine activations. Therefore, the only structure systematically shared between the two sets is the underlying static cornerstone subspace, $V_C$. The adversary recovers a basis for $V_C$ by computing the intersection of the vector spaces spanned by set $A$ and set $B$ (\numcircled{3}). Note that the individual fingerprint vectors within set $A$ will be different from those within set $B$. We can still calculate the intersection subspace because even though these fingerprint vectors are distinct, those within set $A$ and those within set $B$ each span the cornerstone fingerprint subspace $V_C$. Calculating the subspace intersection reduces to solving a system of linear equations (Appendix \ref{app:subspace_intersection}).

\subsubsection{Stage 2: Identifying the fingerprint to bypass the integrity check.}
Once the attacker has uncovered the fingerprint subspace, $V_C$, they use it to bypass the integrity checks as follows:
\begin{itemize}
    \item If a vector lies inside the subspace $V_C$, it is identified as an integrity challenge. To remain undetected, the attacker allows it to be processed correctly, passing the integrity check.
    \item If a vector lies outside the subspace $V_C$, it is identified as a genuine user query. The attacker is now free to return a tampered result to the application \numcircled{6}.
\end{itemize}
This attack fully compromises Soter's integrity mechanism, demonstrating that the efficiency-driven static basis design pattern is insecure even in sophisticated protocols. 

\section{Evaluation}
\label{sec:evaluation}
In this section, we empirically validate the feasibility and effectiveness of our proposed attacks. Our evaluation is structured to answer six key questions:

\begin{tcolorbox}[colback=black!5!white,colframe=black!75!black, left=3pt, right=3pt,top=0pt,bottom=0pt]
{\small
    \textbf{RQ1:} Why is the vulnerable precomputation pattern a necessary choice for designers? (\S\ref{sec:eval:performance}) \\
    \textbf{RQ2:} Can an attacker discover the secret noise dimension, $K$, to get a foothold? (\S\ref{sec:eval:k_discovery}) \\
    \textbf{RQ3:} How effective are the attacks once the prerequisite knowledge is obtained? (\S\ref{sec:eval:efficacy}) \\
    \textbf{RQ4:} Are the attacks robust against simple countermeasures like random subset sampling or increasing $K$? (\S\ref{sec:eval:robustness}) \\
    \textbf{RQ5:} Are the attacks practical and scalable against large, state-of-the-art models? (\S\ref{sec:eval:scalability}) \\
    \textbf{RQ6:} What is the final impact of a successful attack on the model's IP? (\S\ref{sec:eval:impact})
    }
\end{tcolorbox}

\subsection{Experimental Setup}\label{subsec:exp-setup}
\paragraph{Setup.} Our experiments are conducted on a server with a dual-socket Intel Xeon Silver 4309Y CPU (2.80 GHz, SGXv2 enabled with a 4 GB EPC, of which 512 MB is allocated as heap space for our enclave) and a colocated NVIDIA A100-80GB GPU. The software stack consists of Rocky Linux 9.5, Intel SGX SDK v2.25, and PyTorch 2.5.1.

\paragraph{Implementation.}
Our implementation consists of two components. First, for the real-world performance benchmarks (\textbf{RQ1, RQ6}), we built a lightweight, single-threaded C++ library to execute the TEE-based cryptographic protocols on our physical SGX hardware. This library implements standard optimizations like fast modular reduction and memory tiling. Second, for a controlled analysis of the protocol-level vulnerabilities (\textbf{RQ2--RQ6}), we implemented faithful simulations of the core protocols from TLG and Soter. All finite field $\mathbb{F}_P$ computations are done with the prime $P = 2^{31}-1$ unless stated otherwise. For the integrity attack simulation, we assume activations are drawn from a normal distribution.

\subsection{The Performance Imperative for Precomputation}
\label{sec:eval:performance}
Our central claim is that the static basis vulnerability arises from a severe security-performance trade-off. To validate this premise, we first benchmark the performance of a TEE-shielded LLaMA-3 8B model. We implement the TEE-based model confidentiality protocol presented in TLG, as detailed in Section~\ref{sec:bg:tee-based-auth}. To compute the noise, $m$, and its noise effect, $mW_3$ (\numcircled{6} and \numcircled{8} in Figure~\ref{fig:onthefly_vs_precomputation}), we compare two cryptographic strategies analyzed in Section~\ref{sec:onthefly_vs_precomputation}: a secure on-the-fly method that generates fresh randomness for each operation, and the precomputation method that uses a static basis.

\begin{figure}[t]
    \centering
    \includegraphics[width=\columnwidth]{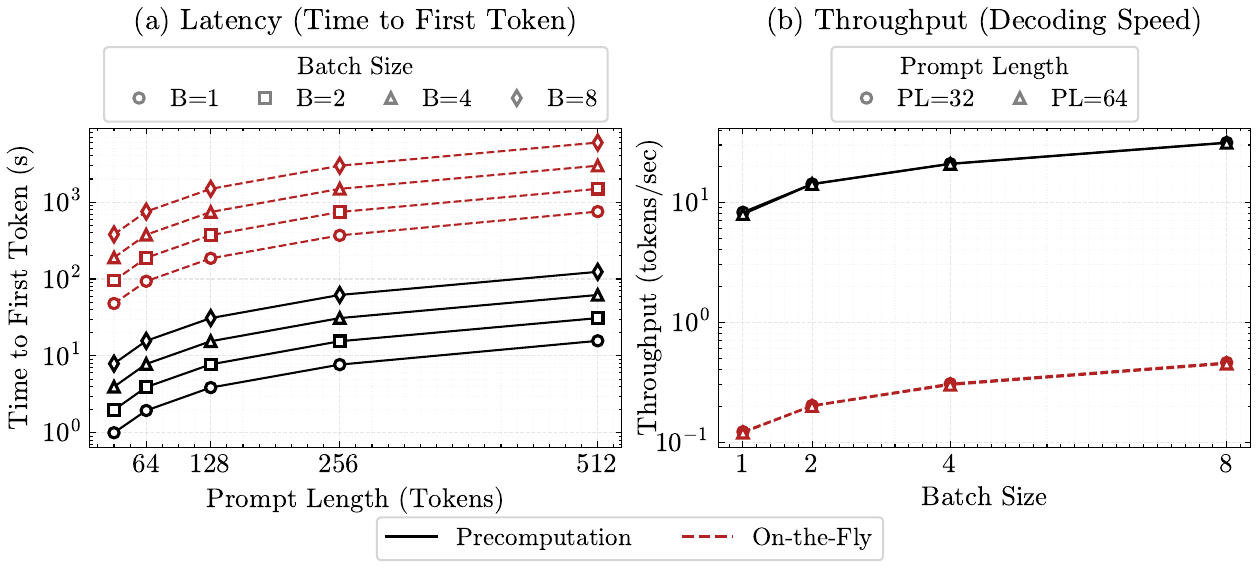}
    \caption{Performance comparison demonstrating the security-performance tension. The secure on-the-fly mitigation exhibits (a) orders-of-magnitude higher latency and (b) lower throughput than the vulnerable precomputation system, rendering it impractical for real-world use.
    \vspace{-.2in}
    }
    \label{fig:mitigation_otf_vs_precomputation}
\end{figure}

As shown in Figure~\ref{fig:mitigation_otf_vs_precomputation}, the secure on-the-fly noise generation method is 37x - 55x slower than the precomputation strategy. For a typical interactive scenario (batch size 1, 128-token prompt), the on-the-fly method has a time-to-first-token (TTFT) of 187 seconds, compared to 3.61 seconds for the precomputation strategy. This is a 52x slowdown, making it unsuitable for real-world applications. 
The impact on throughput is also severe. At a batch size of 8, the on-the-fly method achieves only 0.5 tokens/sec (TPS), while the precomputation strategy delivers 18.6 TPS. This prohibitive performance gap makes the secure on-the-fly method impractical for any system requiring low-latency inference.

\begin{tcolorbox}[colback=black!5!white,colframe=black!75!black,left=3pt,right=3pt,top=0pt,bottom=0pt]
{\small
    \textbf{Answer to RQ1:} The secure on-the-fly method imposes prohibitive performance costs on TTFT and TPS. This effectively pushes designers to use the faster but vulnerable precomputation approach.
    }
\end{tcolorbox}

\subsection{Discoverability of Secret Noise Dimension $K$}
\label{sec:eval:k_discovery}
A key artifact of the static basis vulnerability is that the secret noise dimension, K, is empirically observable. While our threat model grants the attacker knowledge of this parameter, we first demonstrate it can be discovered empirically. This initial step showcases how the core algebraic principles of our attack serve as a versatile method to snoop for information leakage inherent to the precomputation-based pattern. We show this for both of our target systems.

\begin{figure}[t]
    \centering
    \includegraphics[width=\columnwidth]{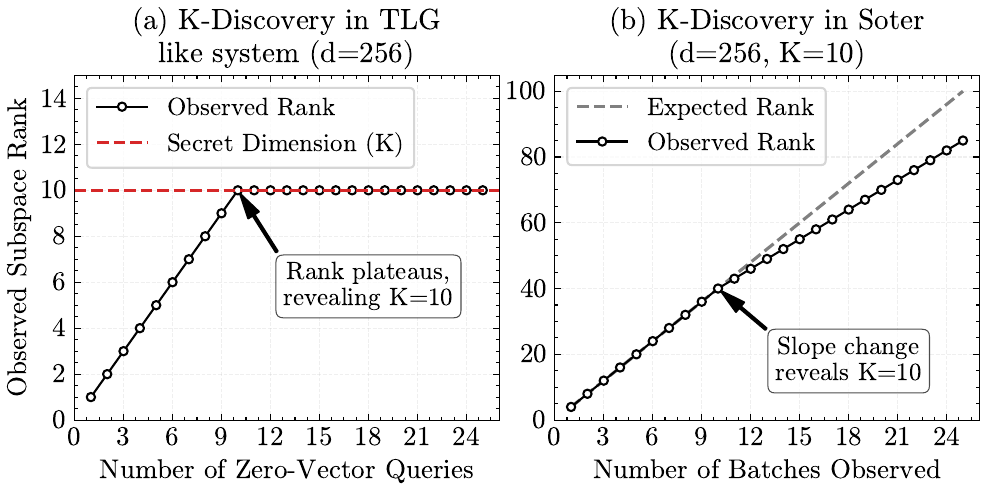}
    \caption{Empirical discovery of the secret noise dimension K. 
    \textbf{(a)} In the direct system, the rank of observed outputs stabilizes at K=10. 
    \textbf{(b)} In the hidden system (Soter), a distinct change in the slope of rank growth reveals K=10.
    \vspace{-.2in}
    }
    \label{fig:rank_determination_combined}
\end{figure}

For a TLG-like system, an attacker can perform the first stage of the attack described in Section~\ref{sec:direct_subspace_attack} by sending zero-vector queries (\numcircled{1} in Figure~\ref{fig:direct_attack}) and observing the rank of the resulting output vectors. As shown in Figure~\ref{fig:rank_determination_combined} (a), the rank grows linearly with each new observation until it stabilizes precisely at $K$, which reveals the system parameter.

For a more complex system like Soter, an attacker can passively execute the first part of the subspace recovery stage of the attack detailed in Section~\ref{sec:cross_attack_soter}. As the attacker observes batches of mixed data (\numcircled{1} in Figure~\ref{fig:intersection_attack}), the rank of the collected vectors initially grows linearly. However, as shown in Figure~\ref{fig:rank_determination_combined}(b), the rank growth soon deviates from the expected trajectory, revealing a distinct change in slope. This occurs because the low-dimensional fingerprints, all drawn from the same precomputed basis, contribute progressively less new information. This deviation allows an attacker to infer the dimension of the fingerprint subspace, $K$.

\begin{tcolorbox}[colback=black!5!white,colframe=black!75!black,left=3pt,right=3pt,top=0pt,bottom=0pt]
{\small
    \textbf{Answer to RQ2:} Yes, $K$ is leaked in both systems. An active attacker can discover $K$ in a TLG-like system by observing the output rank plateauing, similarly for Soter-like systems.
}
\end{tcolorbox}

\subsection{Efficacy: A Deterministic and Total System Break}
\label{sec:eval:efficacy}
Having established that an attacker can discover $K$, we now demonstrate the attack's efficacy. The success of the final attack is critically dependent on the initial stage of learning the complete noise subspace. To show this, we measure the success rate of the full attack (Stage 1 + Stage 2) while varying the number of queries used to learn the subspace (Stage 1).

\begin{figure}[t]
    \centering
    \includegraphics[width=\columnwidth]{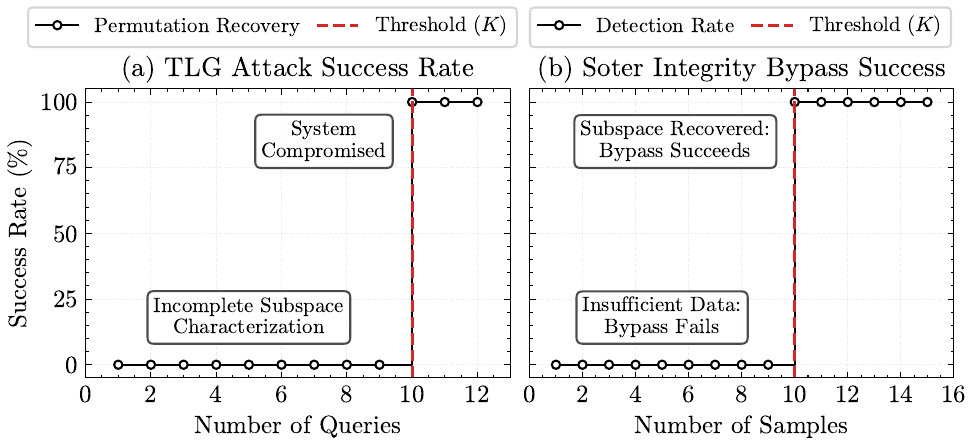}
    \caption{This figure plots the success rate of the full attack (Stage 2) as a function of the number of queries used in the initial subspace learning stage (Stage 1). The attack is completely ineffective if the noise subspace is not fully characterized. Once Stage 1 is completed with at least $K$ queries, the final success rate for a full compromise jumps to 100\%.
    \vspace{-.2in}
    }
    \label{fig:efficacy_combined}
\end{figure}

The results, shown in Figure~\ref{fig:efficacy_combined}, reveal a sharp threshold. For our attack on a TLG-like system (a), the final permutation recovery is ineffective when the noise subspace is characterized with fewer than $K$ queries (\numcircled{1} in Figure~\ref{fig:direct_attack}). However, once the subspace is learned with exactly $K$ queries, the success rate of the final attack stage jumps to 100\%. This demonstrates that fully characterizing the noise subspace is the essential prerequisite for a successful compromise. A similar deterministic threshold is observed for our attack on Soter's integrity mechanism (b). By observing $K$ samples for each of its two sets (\numcircled{1} in Figure~\ref{fig:intersection_attack}), an attacker can fully recover the cornerstone subspace and reliably distinguish genuine activations from integrity checks.

\begin{tcolorbox}[colback=black!5!white,colframe=black!75!black,left=3pt,right=3pt,top=0pt,bottom=0pt]
{\small
    \textbf{Answer to RQ3:} The attacks are 100\% effective. They exhibit a sharp, deterministic threshold, transitioning from 0\% to 100\% success for a full compromise the moment the noise subspace is fully characterized.
}
\end{tcolorbox}

\subsection{Robustness Against Countermeasures}
\label{sec:eval:robustness}
A robust attack must hold up against plausible defenses. We test our attacks against two intuitive countermeasures: (1) using a more complex random subset sampling to generate noise, and (2) increasing the size of the static basis, $K$.

\begin{figure}[t]
\centering
\includegraphics[width=\columnwidth]{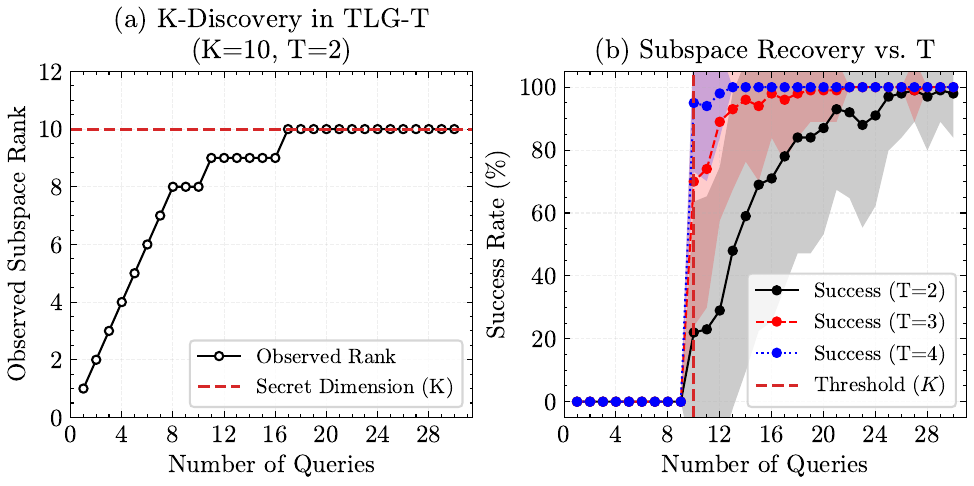}
\caption{Our attack on TLG's model confidentiality protocol remains effective against a sampling strategy that first chooses $T$ out of $K$ noise vectors. (a) An attacker can still discover the secret dimension $K$ by observing the rank plateau of noise samples. (b) Although smaller $T$ values require more samples, the attacker can still achieve a $100\%$ success rate in recovering the full noise subspace.}
\label{fig:direct_attack_t_analysis}
\end{figure}

\begin{figure}[t]
\centering
\includegraphics[width=\columnwidth]{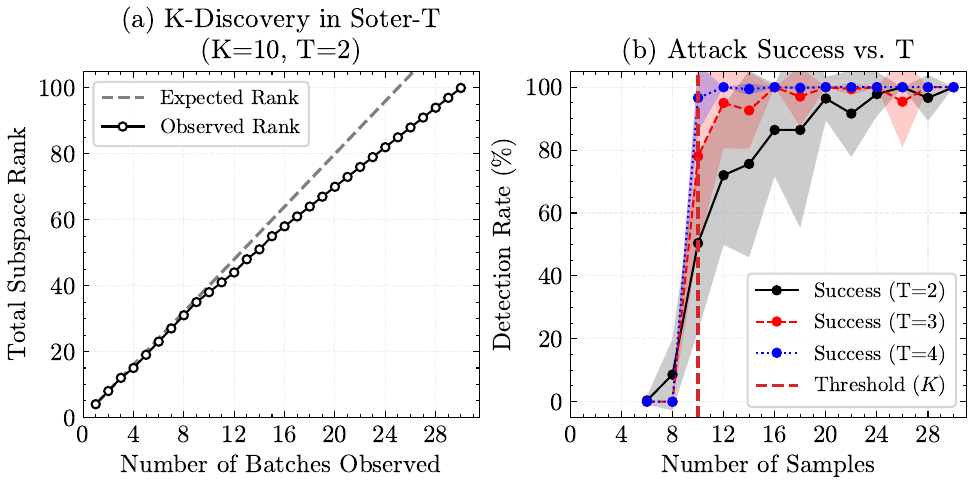}
\caption{Our attack on Soter's integrity checking mechanism remains effective even against different sampling strategies. (a) The secret dimension $K$ is still discoverable from rank growth deviation, and (b) while requiring more samples for smaller $T$, the attack reliably converges to a 100\% success rate, proving the defense is ineffective.
\vspace{-.2in}
}
\label{fig:soter_t_analysis}
\end{figure}

First, we evaluate the random subset sampling strategy, where for each query, the TEE generates noise from a random subset of $T$ basis vectors ($T < K$). As shown in Figures~\ref{fig:direct_attack_t_analysis} and \ref{fig:soter_t_analysis}, this defense is ineffective. The secret dimension $K$ remains discoverable (Panel (a) in both figures), and the attack's success rate still reliably converges to 100\% (Panel (b) in both figures), albeit with more samples.

\begin{figure}[t]
    \centering
    \includegraphics[width=\columnwidth]{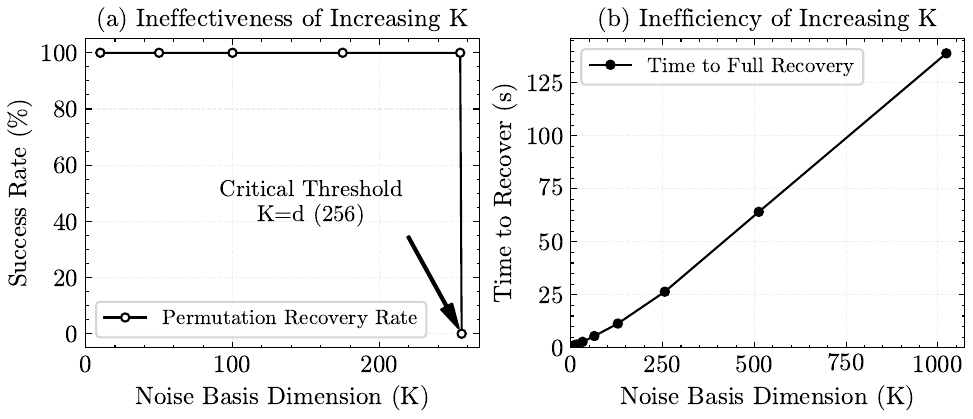}
    \caption{Analysis of increasing the noise basis size (K) as a defense against the attack on TLG's model confidentiality protocol. (a) The attack success rate remains at 100\%, proving that increasing K is an ineffective countermeasure. (b) The time-to-attack scales linearly, showing that this defense only predictably delays the compromise.
    \vspace{-.2in}
    }
    \label{fig:scalability_vs_k}
\end{figure}

\begin{figure}[t]
    \centering
    \includegraphics[width=\columnwidth]{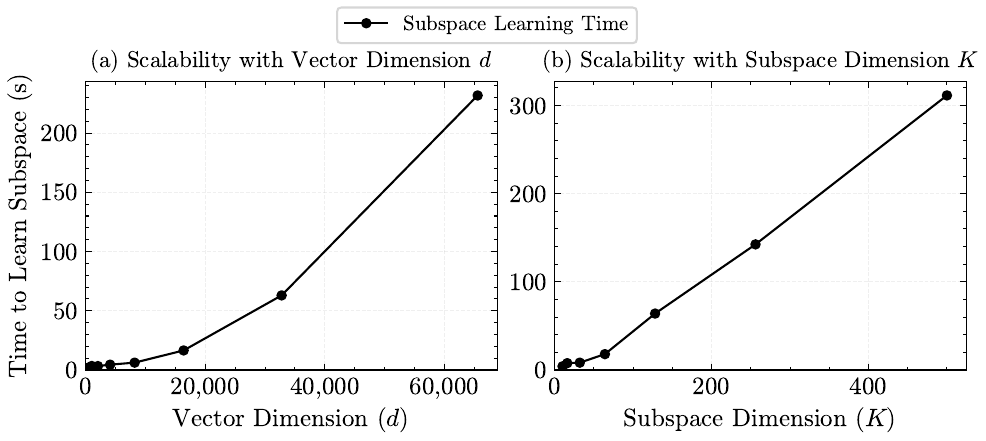}
    \caption{Scalability of the attack on Soter's integrity checking mechanism. The time required to learn the secret subspace scales predictably with both (a) the vector dimension $d$ and (b) the subspace dimension $K$.
    \vspace{-.2in}
    }
    \label{fig:soter_scalability}
\end{figure}

Second, we evaluate the strategy of increasing the basis size $K$. Figures~\ref{fig:scalability_vs_k} and \ref{fig:soter_scalability}(b) show this is also futile. The attack success rate remains at 100\% regardless of $K$. The only effect is that the time-to-compromise scales linearly and predictably with $K$. 

Both experiments demonstrate that these countermeasures do not fix the fundamental flaw. Increasing $K$ only makes the attack more expensive, and subset sampling only makes it less efficient, but neither prevents the break.

\begin{tcolorbox}[colback=black!5!white,colframe=black!75!black,left=3pt,right=3pt,top=0pt,bottom=0pt]
{\small
    \textbf{Answer to RQ4:} The attacks are robust. Countermeasures like random subset sampling or increasing $K$ are ineffective. 
}
\end{tcolorbox}


\subsection{Scalability and Practicality: A Real-World Threat}
\label{sec:eval:scalability}

A viable attack must be feasible against large, real-world targets. Here, we analyze the scalability of our attacks to demonstrate their practicality. The primary evidence of practicality is the time-to-compromise. Table~\ref{tab:attack_feasibility_dimension} details the time required to recover a single layer's secret permutation for several state-of-the-art LLMs secured using TLG with a precomputation-based approach. The results show that the threat is highly practical: for a model like LLaMA-3 8B, this process takes approximately 6.2 minutes. Even for a massive 405B parameter model, a single layer's secrets can be recovered in under two hours. These timings were achieved with a single-threaded proof-of-concept implementation. As the attack's core linear algebra is inherently parallelizable~\cite{golub_matrix_1996, blackford_scalapack_1996}, a dedicated attacker could be significantly faster.

Furthermore, our integrity attack on Soter exhibits similarly manageable scaling. As shown in Figure~\ref{fig:soter_scalability}(a), the time required to learn the integrity secret subspace also scales predictably and polynomially with the model's vector dimension, $d$. This reinforces that the underlying vulnerability is consistently exploitable at scale across different systems.

\begin{tcolorbox}[colback=black!5!white,colframe=black!75!black,left=3pt,right=3pt,top=0pt,bottom=0pt]
{\small
    \textbf{Answer to RQ5:} Yes, the attacks are highly practical. The confidentiality attack can recover a full layer's secrets in approximately 6.2 minutes for a LLaMA-3 8B model and in under 101 minutes for a 405B model. The integrity attack shows similarly predictable, polynomial scaling.
}
\end{tcolorbox}

\begin{table}[h]
\centering
\resizebox{\columnwidth}{!}{%
\begin{tabular}{@{}lrrr@{}}
\toprule
\textbf{Model} & \textbf{Dimension ($d$)} & \textbf{Required Queries} & \textbf{Time-to-Attack (min)} \\ \midrule
LLaMA-3 8B & 14336 & 14346 & $\approx 6.2$ \\
Gemma 3 27B & 21504 & 21514 & $\approx 14.3$ \\
LLaMA-3 70B & 28672 & 28682 & $\approx 25.4$ \\
LLAMA-3.1 405B & 53248 & 53258 & $\approx 100.7$ \\
\bottomrule
\end{tabular}%
} 
\caption{Attack feasibility across models (K=10).
\vspace{-.2in}
}
\label{tab:attack_feasibility_dimension}
\end{table}

\subsection{Final Impact: Total IP and Integrity Compromise}
\label{sec:eval:impact}

The final impact of our attacks is a versatile and total compromise of the target system's security. For the confidentiality attack, Table~\ref{tab:reconstructed_results} shows the attacker steals a perfect replica of the secret model, achieving identical accuracy and restoring full native performance. For the integrity attack, successfully recovering the fingerprinting subspace allows an attacker to tamper with any offloaded computation without detection. In both cases, the TEE-based protections are rendered completely ineffective due to the vulnerability arising from the use of precomputed noise.

\begin{tcolorbox}[colback=black!5!white,colframe=black!75!black,left=3pt,right=3pt,top=2pt,bottom=2pt]
{\small
    \textbf{Answer to RQ6:} The final impact is a total compromise of both security guarantees. The confidentiality attack yields an exact replica of the secret model, and the integrity attack allows undetected tampering with any offloaded computation.
}
\end{tcolorbox}

\begin{table}[h]
\centering
\resizebox{\columnwidth}{!}{%
\begin{tabular}{@{}lrrr@{}}
\toprule
\textbf{System Configuration} & \textbf{GSM8K Accuracy (\%)} & \textbf{Throughput (tok/s)} & \textbf{TEE Required?} \\ \midrule
Unprotected Baseline (LLaMA-3 8B) & 82.34 & 240 & No \\
Original TEE-Shielded System & 82.34 & 18 & Yes \\
\textbf{Attacker-Reconstructed Model} & \textbf{82.34} & \textbf{240} & \textbf{No} \\ \bottomrule
\end{tabular}%
} 
\caption{Post-attack fidelity and performance. The attacker's reconstructed model achieves identical accuracy and performance to the original unprotected baseline, proving a total compromise of the model's intellectual property.}
\label{tab:reconstructed_results}
\end{table}

\vspace{-.2in}
\section{Related Work}\label{sec:related}

\subsection{TEE-Shielded Model Protection}
Trusted Execution Environments (TEEs) are a hardware-based solution for protecting proprietary machine learning (ML) models on third-party devices. Early solutions in this area focused on executing the entire model within a secure enclave \cite{hanzlik_mlcapsule_2019, ohrimenko_oblivious_2016, kunkel_tensorscone_2019, bayerl_offline_2020}. While this provides strong security, traditional CPU-based TEEs face severe memory and performance constraints that make this impractical for large models \cite{lee_occlumency_2019, kim_vessels_2020, vannostrand_confidential_2019, gangal_hybridtee_2020, islam_confidential_2023, li_secure_2023, hu_secure_2023, quoc_securetf_2020}. Newer confidential computing architectures on GPUs and CPUs, such as NVIDIA Hopper Confidential Computing and Arm's CCA, are beginning to address these limitations by enabling the secure execution of entire large models \cite{abdollahi_early_2025, siby_guarantee_2024, dong_evaluating_2025, zhu_confidential_2024}.

To leverage powerful but untrusted hardware accelerators like GPUs, Partial TEE-Shielded Execution (PTSE) approaches have been introduced. A crude PTSE strategy involves model partitioning, where the model is split and only the most sensitive layers are executed within the TEE \cite{mo_darknetz_2020, schlogl_ennclave_2020, liu_mirrornet_2023, liu_tbnet_2024, narra_privacy-preserving_2019, mo_machine_2024}. Sophisticated PTSE strategies involve cryptographic obfuscation \cite{tramer_slalom_2019, hou_model_2022, bruhns_slalom_2024, hashemi_darknight_2022, xu_tempo_2024, sun_tensorshield_2025, bai_phantom_2025}, where weights are decomposed \cite{refael_slip_2024}, transformed, masked, or permuted inside the TEE before being offloaded for computation \cite{shen_soter_2022, sun_shadownet_2023, li_translinkguard_2024}. The performance demands of modern LLMs mean these obfuscation schemes either explicitly adopt a precomputed, static secret basis to accelerate cryptographic operations \cite{shen_soter_2022, sun_tsqp_2025, xu_tempo_2024}, or would require such a design to scale efficiently, making them susceptible to the same vulnerability \cite{zhang_groupcover_2024, sun_shadownet_2023, li_translinkguard_2024, zhang_no_2023, wang_game_nodate, li_teeslice_2024}. Our work is the first to formally identify this performance-driven design pattern as a fundamental flaw and demonstrate its exploitability through a deterministic, cross-query algebraic attack.

\subsection{Attack Landscape for Secure Inference}
\subsubsection*{Orthogonal Attack Vectors}
A significant body of work targets secure inference through vectors orthogonal to our protocol-level analysis. Side-channel attacks exploit hardware-level information leakage like cache timing~\cite{bernstein_cache-timing_2005, kocher_timing_1996}, memory access patterns~\cite{yarom_flushreload_2014}, or power consumption~\cite{kocher_differential_1999} to infer secrets, and have been used to leak user inputs \cite{yang_first_2024, sorensen_leftoverlocals_2024, adiletta_spill_2025, yuan_ciphersteal_2024, banerjee_triton_2023, grover_privado_2019} and model secrets \cite{hua_reverse_2018, yan_cache_2020, wei_leaky_2020, gao_deeptheft_2024, rakin_deepsteal_2022}. Meanwhile, model extraction aims to steal a model's functionality by training a substitute on its API outputs \cite{tramer_stealing_2016, sanyal_towards_2022, rolnick_reverse-engineering_2019, oliynyk_i_2025, nayan_sok_nodate, jagielski_high_2020, finlayson_logits_2024, carlini_cryptanalytic_2020, carlini_stealing_2024}. Our attack is fundamentally different: we assume a perfect TEE free from side channels and exploit a mathematical flaw in the security protocol itself, rather than attacking its physical implementation or stealing its learned function.

\subsubsection*{Protocol-Level Flaws}
Prior work has also identified protocol-level flaws, but these typically exploit statistical properties of the obfuscated weights to infer shielded components by matching them against a known public model \cite{zhang_groupcover_2024, wang_game_nodate, zhang_no_2023}. Consequently, the effectiveness of this class of attacks is limited, as they are predicated on the assumption that the victim model is fine-tuned from a public model. Our attack is fundamentally different. It is algebraic and deterministic; we do not infer secrets from noisy correlations but construct a system of linear equations from multiple queries to solve for the exact secret basis. Our contribution is thus more closely aligned with the well-understood category of vulnerabilities stemming from the reuse of secret keying material \cite{stinson_cryptography_2018, vanhoef_key_2017, schneier_cryptanalysis_1998, bock_nonce-disrespecting_2016}. We identify the static secret basis pattern as a new manifestation of this flaw within TEE-shielded ML and are the first to demonstrate how it enables an adversary to aggregate information across queries to algebraically break the system's security.

\section{Conclusion}
\label{sec:conclusion}
This work identifies the use of a precomputed static basis, an appealing optimization that performance-bound TEE-shielded systems are often forced to adopt, as a fundamental vulnerability. We prove its severity through two distinct algebraic attacks that defeat both confidentiality and integrity guarantees: one fully recovers secret permutations in a TLG-like system, while the other completely bypasses the integrity checks in Soter. Our findings reveal a critical tension between this optimization and provable security, highlighting the need for new protocols that achieve high performance without reusing secret material.

\bibliographystyle{ACM-Reference-Format}
\bibliography{references}


\begin{thebibliography}{96}


\ifx \showCODEN    \undefined \def \showCODEN     #1{\unskip}     \fi
\ifx \showDOI      \undefined \def \showDOI       #1{#1}\fi
\ifx \showISBNx    \undefined \def \showISBNx     #1{\unskip}     \fi
\ifx \showISBNxiii \undefined \def \showISBNxiii  #1{\unskip}     \fi
\ifx \showISSN     \undefined \def \showISSN      #1{\unskip}     \fi
\ifx \showLCCN     \undefined \def \showLCCN      #1{\unskip}     \fi
\ifx \shownote     \undefined \def \shownote      #1{#1}          \fi
\ifx \showarticletitle \undefined \def \showarticletitle #1{#1}   \fi
\ifx \showURL      \undefined \def \showURL       {\relax}        \fi
\providecommand\bibfield[2]{#2}
\providecommand\bibinfo[2]{#2}
\providecommand\natexlab[1]{#1}
\providecommand\showeprint[2][]{arXiv:#2}

\bibitem[Abdollahi et~al\mbox{.}(2025)]%
        {abdollahi_early_2025}
\bibfield{author}{\bibinfo{person}{Sina Abdollahi}, \bibinfo{person}{Mohammad
  Maheri}, \bibinfo{person}{Sandra Siby}, \bibinfo{person}{Marios Kogias},
  {and} \bibinfo{person}{Hamed Haddadi}.} \bibinfo{year}{2025}\natexlab{}.
\newblock \bibinfo{title}{An {Early} {Experience} with {Confidential}
  {Computing} {Architecture} for {On}-{Device} {Model} {Protection}}.
\newblock
\newblock
\urldef\tempurl%
\url{https://doi.org/10.48550/arXiv.2504.08508}
\showDOI{\tempurl}


\bibitem[Adiletta and Sunar(2025)]%
        {adiletta_spill_2025}
\bibfield{author}{\bibinfo{person}{Andrew Adiletta} {and} \bibinfo{person}{Berk
  Sunar}.} \bibinfo{year}{2025}\natexlab{}.
\newblock \bibinfo{title}{Spill {The} {Beans}: {Exploiting} {CPU} {Cache}
  {Side}-{Channels} to {Leak} {Tokens} from {Large} {Language} {Models}}.
\newblock
\newblock
\urldef\tempurl%
\url{https://doi.org/10.48550/arXiv.2505.00817}
\showDOI{\tempurl}


\bibitem[Anthropic(2025)]%
        {anthropic_meet_nodate}
\bibfield{author}{\bibinfo{person}{Anthropic}.}
  \bibinfo{year}{2025}\natexlab{}.
\newblock \bibinfo{title}{Meet {Claude} {Anthropic}}.
\newblock
\newblock
\urldef\tempurl%
\url{https://www.anthropic.com/claude}
\showURL{%
\tempurl}


\bibitem[Apple(2025)]%
        {apple_deploying_nodate}
\bibfield{author}{\bibinfo{person}{Apple}.} \bibinfo{year}{2025}\natexlab{}.
\newblock \bibinfo{title}{Deploying {Transformers} on the {Apple} {Neural}
  {Engine}}.
\newblock
\newblock
\urldef\tempurl%
\url{https://machinelearning.apple.com/research/neural-engine-transformers}
\showURL{%
\tempurl}


\bibitem[Bai et~al\mbox{.}(2025)]%
        {bai_phantom_2025}
\bibfield{author}{\bibinfo{person}{Juyang Bai},
  \bibinfo{person}{Md~Hafizul~Islam Chowdhuryy}, \bibinfo{person}{Jingtao Li},
  \bibinfo{person}{Fan Yao}, \bibinfo{person}{Chaitali Chakrabarti}, {and}
  \bibinfo{person}{Deliang Fan}.} \bibinfo{year}{2025}\natexlab{}.
\newblock \showarticletitle{Phantom: {Privacy}-{Preserving} {Deep} {Neural}
  {Network} {Model} {Obfuscation} in {Heterogeneous} {TEE} and {GPU} {System}}.
  \bibinfo{pages}{5565--5582}.
\newblock
\showISBNx{978-1-939133-52-6}
\urldef\tempurl%
\url{https://www.usenix.org/conference/usenixsecurity25/presentation/bai-juyang}
\showURL{%
\tempurl}


\bibitem[Banerjee et~al\mbox{.}(2023)]%
        {banerjee_triton_2023}
\bibfield{author}{\bibinfo{person}{Sarbartha Banerjee}, \bibinfo{person}{Shijia
  Wei}, \bibinfo{person}{Prakash Ramrakhyani}, {and} \bibinfo{person}{Mohit
  Tiwari}.} \bibinfo{year}{2023}\natexlab{}.
\newblock \showarticletitle{Triton: {Software}-{Defined} {Threat} {Model} for
  {Secure} {Multi}-{Tenant} {ML} {Inference} {Accelerators}}. In
  \bibinfo{booktitle}{\emph{Proceedings of the 12th {International} {Workshop}
  on {Hardware} and {Architectural} {Support} for {Security} and {Privacy}}}
  \emph{(\bibinfo{series}{{HASP} '23})}. \bibinfo{publisher}{Association for
  Computing Machinery}, \bibinfo{address}{New York, NY, USA},
  \bibinfo{pages}{19--28}.
\newblock
\showISBNx{979-8-4007-1623-2}
\urldef\tempurl%
\url{https://doi.org/10.1145/3623652.3623672}
\showDOI{\tempurl}


\bibitem[Bayerl et~al\mbox{.}(2020)]%
        {bayerl_offline_2020}
\bibfield{author}{\bibinfo{person}{Sebastian~P. Bayerl},
  \bibinfo{person}{Tommaso Frassetto}, \bibinfo{person}{Patrick Jauernig},
  \bibinfo{person}{Korbinian Riedhammer}, \bibinfo{person}{Ahmad-Reza Sadeghi},
  \bibinfo{person}{Thomas Schneider}, \bibinfo{person}{Emmanuel Stapf}, {and}
  \bibinfo{person}{Christian Weinert}.} \bibinfo{year}{2020}\natexlab{}.
\newblock \showarticletitle{Offline {Model} {Guard}: {Secure} and {Private}
  {ML} on {Mobile} {Devices}}. In \bibinfo{booktitle}{\emph{2020 {Design},
  {Automation} \& {Test} in {Europe} {Conference} \& {Exhibition} ({DATE})}}.
  \bibinfo{pages}{460--465}.
\newblock
\urldef\tempurl%
\url{https://doi.org/10.23919/DATE48585.2020.9116560}
\showDOI{\tempurl}


\bibitem[Bernstein(2005)]%
        {bernstein_cache-timing_2005}
\bibfield{author}{\bibinfo{person}{D. Bernstein}.}
  \bibinfo{year}{2005}\natexlab{}.
\newblock \showarticletitle{Cache-timing attacks on {AES}}.
\newblock
\urldef\tempurl%
\url{https://www.semanticscholar.org/paper/Cache-timing-attacks-on-AES-Bernstein/352e74019d86163d73618f03429ae452ab429629}
\showURL{%
\tempurl}


\bibitem[Blackford et~al\mbox{.}(1996)]%
        {blackford_scalapack_1996}
\bibfield{author}{\bibinfo{person}{Laura~Susan Blackford}, \bibinfo{person}{J.
  Choi}, \bibinfo{person}{A. Cleary}, \bibinfo{person}{A. Petitet},
  \bibinfo{person}{R.~C. Whaley}, \bibinfo{person}{J. Demmel},
  \bibinfo{person}{I. Dhillon}, \bibinfo{person}{K. Stanley},
  \bibinfo{person}{J. Dongarra}, \bibinfo{person}{S. Hammarling},
  \bibinfo{person}{G. Henry}, {and} \bibinfo{person}{D. Walker}.}
  \bibinfo{year}{1996}\natexlab{}.
\newblock \showarticletitle{{ScaLAPACK}: a portable linear algebra library for
  distributed memory computers - design issues and performance}. In
  \bibinfo{booktitle}{\emph{Proceedings of the 1996 {ACM}/{IEEE} conference on
  {Supercomputing}}} \emph{(\bibinfo{series}{Supercomputing '96})}.
  \bibinfo{publisher}{IEEE Computer Society}, \bibinfo{address}{USA},
  \bibinfo{pages}{5--es}.
\newblock
\showISBNx{978-0-89791-854-1}
\urldef\tempurl%
\url{https://doi.org/10.1145/369028.369038}
\showDOI{\tempurl}


\bibitem[Bruhns et~al\mbox{.}(2024)]%
        {bruhns_slalom_2024}
\bibfield{author}{\bibinfo{person}{Ida Bruhns}, \bibinfo{person}{Sebastian
  Berndt}, \bibinfo{person}{Jonas Sander}, {and} \bibinfo{person}{Thomas
  Eisenbarth}.} \bibinfo{year}{2024}\natexlab{}.
\newblock \showarticletitle{Slalom at the {Carnival}: {Privacy}-preserving
  {Inference} with {Masks} from {Public} {Knowledge}}.
\newblock \bibinfo{journal}{\emph{IACR Communications in Cryptology}}
  \bibinfo{volume}{1}, \bibinfo{number}{3} (\bibinfo{date}{Oct.}
  \bibinfo{year}{2024}).
\newblock
\showISSN{3006-5496}
\urldef\tempurl%
\url{https://doi.org/10.62056/akp-49qgxq}
\showDOI{\tempurl}


\bibitem[Böck et~al\mbox{.}(2016)]%
        {bock_nonce-disrespecting_2016}
\bibfield{author}{\bibinfo{person}{Hanno Böck}, \bibinfo{person}{Aaron
  Zauner}, \bibinfo{person}{Sean Devlin}, \bibinfo{person}{Juraj Somorovsky},
  {and} \bibinfo{person}{Philipp Jovanovic}.} \bibinfo{year}{2016}\natexlab{}.
\newblock \showarticletitle{Nonce-{Disrespecting} {Adversaries}: {Practical}
  {Forgery} {Attacks} on {GCM} in {TLS}}.
\newblock
\urldef\tempurl%
\url{https://www.usenix.org/conference/woot16/workshop-program/presentation/bock}
\showURL{%
\tempurl}


\bibitem[Carlini et~al\mbox{.}(2020)]%
        {carlini_cryptanalytic_2020}
\bibfield{author}{\bibinfo{person}{Nicholas Carlini}, \bibinfo{person}{Matthew
  Jagielski}, {and} \bibinfo{person}{Ilya Mironov}.}
  \bibinfo{year}{2020}\natexlab{}.
\newblock \bibinfo{title}{Cryptanalytic {Extraction} of {Neural} {Network}
  {Models}}.
\newblock
\newblock
\urldef\tempurl%
\url{https://doi.org/10.48550/arXiv.2003.04884}
\showDOI{\tempurl}


\bibitem[Carlini et~al\mbox{.}(2024)]%
        {carlini_stealing_2024}
\bibfield{author}{\bibinfo{person}{Nicholas Carlini}, \bibinfo{person}{Daniel
  Paleka}, \bibinfo{person}{Krishnamurthy~Dj Dvijotham},
  \bibinfo{person}{Thomas Steinke}, \bibinfo{person}{Jonathan Hayase},
  \bibinfo{person}{A.~Feder Cooper}, \bibinfo{person}{Katherine Lee},
  \bibinfo{person}{Matthew Jagielski}, \bibinfo{person}{Milad Nasr},
  \bibinfo{person}{Arthur Conmy}, \bibinfo{person}{Itay Yona},
  \bibinfo{person}{Eric Wallace}, \bibinfo{person}{David Rolnick}, {and}
  \bibinfo{person}{Florian Tramèr}.} \bibinfo{year}{2024}\natexlab{}.
\newblock \bibinfo{title}{Stealing {Part} of a {Production} {Language}
  {Model}}.
\newblock
\newblock
\urldef\tempurl%
\url{https://doi.org/10.48550/arXiv.2403.06634}
\showDOI{\tempurl}


\bibitem[Cheng et~al\mbox{.}(2025)]%
        {cheng_reclaiming_2025}
\bibfield{author}{\bibinfo{person}{Zerui Cheng}, \bibinfo{person}{Edoardo
  Contente}, \bibinfo{person}{Ben Finch}, \bibinfo{person}{Oleg Golev},
  \bibinfo{person}{Jonathan Hayase}, \bibinfo{person}{Andrew Miller},
  \bibinfo{person}{Niusha Moshrefi}, \bibinfo{person}{Anshul Nasery},
  \bibinfo{person}{Sandeep Nailwal}, \bibinfo{person}{Sewoong Oh},
  \bibinfo{person}{Himanshu Tyagi}, {and} \bibinfo{person}{Pramod Viswanath}.}
  \bibinfo{year}{2025}\natexlab{}.
\newblock \bibinfo{title}{Reclaiming "{Open} {AI}" -- {AI} {Model} {Serving}
  {Can} {Be} {Open} {Access}, {Yet} {Monetizable} and {Loyal}}.
\newblock
\newblock
\urldef\tempurl%
\url{https://doi.org/10.48550/arXiv.2411.03887}
\showDOI{\tempurl}


\bibitem[Chrapek et~al\mbox{.}(2024)]%
        {chrapek_fortify_2024}
\bibfield{author}{\bibinfo{person}{Marcin Chrapek}, \bibinfo{person}{Anjo
  Vahldiek-Oberwagner}, \bibinfo{person}{Marcin Spoczynski},
  \bibinfo{person}{Scott Constable}, \bibinfo{person}{Mona Vij}, {and}
  \bibinfo{person}{Torsten Hoefler}.} \bibinfo{year}{2024}\natexlab{}.
\newblock \bibinfo{title}{Fortify {Your} {Foundations}: {Practical} {Privacy}
  and {Security} for {Foundation} {Model} {Deployments} {In} {The} {Cloud}}.
\newblock
\newblock
\urldef\tempurl%
\url{https://doi.org/10.48550/arXiv.2410.05930}
\showDOI{\tempurl}


\bibitem[Cui et~al\mbox{.}(2024)]%
        {cui_survey_2024}
\bibfield{author}{\bibinfo{person}{Can Cui}, \bibinfo{person}{Yunsheng Ma},
  \bibinfo{person}{Xu Cao}, \bibinfo{person}{Wenqian Ye}, \bibinfo{person}{Yang
  Zhou}, \bibinfo{person}{Kaizhao Liang}, \bibinfo{person}{Jintai Chen},
  \bibinfo{person}{Juanwu Lu}, \bibinfo{person}{Zichong Yang},
  \bibinfo{person}{Kuei-Da Liao}, \bibinfo{person}{Tianren Gao},
  \bibinfo{person}{Erlong Li}, \bibinfo{person}{Kun Tang},
  \bibinfo{person}{Zhipeng Cao}, \bibinfo{person}{Tong Zhou},
  \bibinfo{person}{Ao Liu}, \bibinfo{person}{Xinrui Yan},
  \bibinfo{person}{Shuqi Mei}, \bibinfo{person}{Jianguo Cao},
  \bibinfo{person}{Ziran Wang}, {and} \bibinfo{person}{Chao Zheng}.}
  \bibinfo{year}{2024}\natexlab{}.
\newblock \showarticletitle{A {Survey} on {Multimodal} {Large} {Language}
  {Models} for {Autonomous} {Driving}}. In \bibinfo{booktitle}{\emph{2024
  {IEEE}/{CVF} {Winter} {Conference} on {Applications} of {Computer} {Vision}
  {Workshops} ({WACVW})}}. \bibinfo{pages}{958--979}.
\newblock
\urldef\tempurl%
\url{https://doi.org/10.1109/WACVW60836.2024.00106}
\showDOI{\tempurl}


\bibitem[Dong and Wang(2025)]%
        {dong_evaluating_2025}
\bibfield{author}{\bibinfo{person}{Ben Dong} {and} \bibinfo{person}{Qian
  Wang}.} \bibinfo{year}{2025}\natexlab{}.
\newblock \bibinfo{title}{Evaluating the {Performance} of the {DeepSeek}
  {Model} in {Confidential} {Computing} {Environment}}.
\newblock
\newblock
\urldef\tempurl%
\url{https://doi.org/10.48550/arXiv.2502.11347}
\showDOI{\tempurl}


\bibitem[Finlayson et~al\mbox{.}(2024)]%
        {finlayson_logits_2024}
\bibfield{author}{\bibinfo{person}{Matthew Finlayson}, \bibinfo{person}{Xiang
  Ren}, {and} \bibinfo{person}{Swabha Swayamdipta}.}
  \bibinfo{year}{2024}\natexlab{}.
\newblock \bibinfo{title}{Logits of {API}-{Protected} {LLMs} {Leak}
  {Proprietary} {Information}}.
\newblock
\newblock
\urldef\tempurl%
\url{https://doi.org/10.48550/arXiv.2403.09539}
\showDOI{\tempurl}


\bibitem[Firoozi et~al\mbox{.}(2023)]%
        {firoozi_foundation_2023}
\bibfield{author}{\bibinfo{person}{Roya Firoozi}, \bibinfo{person}{Johnathan
  Tucker}, \bibinfo{person}{Stephen Tian}, \bibinfo{person}{Anirudha Majumdar},
  \bibinfo{person}{Jiankai Sun}, \bibinfo{person}{Weiyu Liu},
  \bibinfo{person}{Yuke Zhu}, \bibinfo{person}{Shuran Song},
  \bibinfo{person}{Ashish Kapoor}, \bibinfo{person}{Karol Hausman},
  \bibinfo{person}{Brian Ichter}, \bibinfo{person}{Danny Driess},
  \bibinfo{person}{Jiajun Wu}, \bibinfo{person}{Cewu Lu}, {and}
  \bibinfo{person}{Mac Schwager}.} \bibinfo{year}{2023}\natexlab{}.
\newblock \bibinfo{title}{Foundation {Models} in {Robotics}: {Applications},
  {Challenges}, and the {Future}}.
\newblock
\newblock
\urldef\tempurl%
\url{https://doi.org/10.48550/arXiv.2312.07843}
\showDOI{\tempurl}


\bibitem[Gangal et~al\mbox{.}(2020)]%
        {gangal_hybridtee_2020}
\bibfield{author}{\bibinfo{person}{Akshay Gangal}, \bibinfo{person}{Mengmei
  Ye}, {and} \bibinfo{person}{Sheng Wei}.} \bibinfo{year}{2020}\natexlab{}.
\newblock \showarticletitle{{HybridTEE}: {Secure} {Mobile} {DNN} {Execution}
  {Using} {Hybrid} {Trusted} {Execution} {Environment}}. In
  \bibinfo{booktitle}{\emph{2020 {Asian} {Hardware} {Oriented} {Security} and
  {Trust} {Symposium} ({AsianHOST})}}. \bibinfo{pages}{1--6}.
\newblock
\urldef\tempurl%
\url{https://doi.org/10.1109/AsianHOST51057.2020.9358260}
\showDOI{\tempurl}


\bibitem[Gao et~al\mbox{.}(2024)]%
        {gao_deeptheft_2024}
\bibfield{author}{\bibinfo{person}{Yansong Gao}, \bibinfo{person}{Huming Qiu},
  \bibinfo{person}{Zhi Zhang}, \bibinfo{person}{Binghui Wang},
  \bibinfo{person}{Hua Ma}, \bibinfo{person}{Alsharif Abuadbba},
  \bibinfo{person}{Minhui Xue}, \bibinfo{person}{Anmin Fu}, {and}
  \bibinfo{person}{Surya Nepal}.} \bibinfo{year}{2024}\natexlab{}.
\newblock \showarticletitle{{DeepTheft}: {Stealing} {DNN} {Model}
  {Architectures} through {Power} {Side} {Channel}}. \bibinfo{publisher}{IEEE
  Computer Society}, \bibinfo{pages}{3311--3326}.
\newblock
\showISBNx{979-8-3503-3130-1}
\urldef\tempurl%
\url{https://doi.org/10.1109/SP54263.2024.00250}
\showDOI{\tempurl}


\bibitem[Gilad-Bachrach et~al\mbox{.}(2016)]%
        {gilad-bachrach_cryptonets_2016}
\bibfield{author}{\bibinfo{person}{Ran Gilad-Bachrach}, \bibinfo{person}{Nathan
  Dowlin}, \bibinfo{person}{Kim Laine}, \bibinfo{person}{Kristin Lauter},
  \bibinfo{person}{Michael Naehrig}, {and} \bibinfo{person}{John Wernsing}.}
  \bibinfo{year}{2016}\natexlab{}.
\newblock \showarticletitle{{CryptoNets}: {Applying} {Neural} {Networks} to
  {Encrypted} {Data} with {High} {Throughput} and {Accuracy}}. In
  \bibinfo{booktitle}{\emph{Proceedings of {The} 33rd {International}
  {Conference} on {Machine} {Learning}}}. \bibinfo{publisher}{PMLR},
  \bibinfo{pages}{201--210}.
\newblock
\urldef\tempurl%
\url{https://proceedings.mlr.press/v48/gilad-bachrach16.html}
\showURL{%
\tempurl}


\bibitem[Golub and Van~Loan(1996)]%
        {golub_matrix_1996}
\bibfield{author}{\bibinfo{person}{Gene~H. Golub} {and}
  \bibinfo{person}{Charles~F. Van~Loan}.} \bibinfo{year}{1996}\natexlab{}.
\newblock \bibinfo{booktitle}{\emph{Matrix computations (3rd ed.)}}.
\newblock \bibinfo{publisher}{Johns Hopkins University Press},
  \bibinfo{address}{USA}.
\newblock
\showISBNx{978-0-8018-5414-9}


\bibitem[Google(2025)]%
        {google_google_nodate}
\bibfield{author}{\bibinfo{person}{Google}.} \bibinfo{year}{2025}\natexlab{}.
\newblock \bibinfo{title}{Google {Gemini}}.
\newblock
\newblock
\urldef\tempurl%
\url{https://gemini.google.com}
\showURL{%
\tempurl}


\bibitem[Grover et~al\mbox{.}(2019)]%
        {grover_privado_2019}
\bibfield{author}{\bibinfo{person}{Karan Grover}, \bibinfo{person}{Shruti
  Tople}, \bibinfo{person}{Shweta Shinde}, \bibinfo{person}{Ranjita Bhagwan},
  {and} \bibinfo{person}{Ramachandran Ramjee}.}
  \bibinfo{year}{2019}\natexlab{}.
\newblock \bibinfo{title}{Privado: {Practical} and {Secure} {DNN} {Inference}
  with {Enclaves}}.
\newblock
\newblock
\urldef\tempurl%
\url{https://doi.org/10.48550/arXiv.1810.00602}
\showDOI{\tempurl}


\bibitem[Gupta(2023)]%
        {gupta_google_2023}
\bibfield{author}{\bibinfo{person}{Monika Gupta}.}
  \bibinfo{year}{2023}\natexlab{}.
\newblock \bibinfo{title}{Google {Tensor} {G3}: {The} new chip that gives your
  {Pixel} an {AI} upgrade}.
\newblock
\newblock
\urldef\tempurl%
\url{https://blog.google/products/pixel/google-tensor-g3-pixel-8/}
\showURL{%
\tempurl}


\bibitem[Hanzlik et~al\mbox{.}(2019)]%
        {hanzlik_mlcapsule_2019}
\bibfield{author}{\bibinfo{person}{Lucjan Hanzlik}, \bibinfo{person}{Yang
  Zhang}, \bibinfo{person}{Kathrin Grosse}, \bibinfo{person}{Ahmed Salem},
  \bibinfo{person}{Max Augustin}, \bibinfo{person}{Michael Backes}, {and}
  \bibinfo{person}{Mario Fritz}.} \bibinfo{year}{2019}\natexlab{}.
\newblock \bibinfo{title}{{MLCapsule}: {Guarded} {Offline} {Deployment} of
  {Machine} {Learning} as a {Service}}.
\newblock
\newblock
\urldef\tempurl%
\url{https://doi.org/10.48550/arXiv.1808.00590}
\showDOI{\tempurl}


\bibitem[Hashemi et~al\mbox{.}(2022)]%
        {hashemi_darknight_2022}
\bibfield{author}{\bibinfo{person}{Hanieh Hashemi}, \bibinfo{person}{Yongqin
  Wang}, {and} \bibinfo{person}{Murali Annavaram}.}
  \bibinfo{year}{2022}\natexlab{}.
\newblock \bibinfo{title}{{DarKnight}: {An} {Accelerated} {Framework} for
  {Privacy} and {Integrity} {Preserving} {Deep} {Learning} {Using} {Trusted}
  {Hardware}}.
\newblock
\newblock
\urldef\tempurl%
\url{https://doi.org/10.48550/arXiv.2207.00083}
\showDOI{\tempurl}


\bibitem[Hong et~al\mbox{.}(2019)]%
        {hong_terminal_2019}
\bibfield{author}{\bibinfo{person}{Sanghyun Hong}, \bibinfo{person}{Pietro
  Frigo}, \bibinfo{person}{Yiğitcan Kaya}, \bibinfo{person}{Cristiano
  Giuffrida}, {and} \bibinfo{person}{Tudor Dumitraș}.}
  \bibinfo{year}{2019}\natexlab{}.
\newblock \showarticletitle{Terminal {Brain} {Damage}: {Exposing} the
  {Graceless} {Degradation} in {Deep} {Neural} {Networks} {Under} {Hardware}
  {Fault} {Attacks}}. \bibinfo{pages}{497--514}.
\newblock
\showISBNx{978-1-939133-06-9}
\urldef\tempurl%
\url{https://www.usenix.org/conference/usenixsecurity19/presentation/hong}
\showURL{%
\tempurl}


\bibitem[Hou et~al\mbox{.}(2022)]%
        {hou_model_2022}
\bibfield{author}{\bibinfo{person}{Jiahui Hou}, \bibinfo{person}{Huiqi Liu},
  \bibinfo{person}{Yunxin Liu}, \bibinfo{person}{Yu Wang},
  \bibinfo{person}{Peng-Jun Wan}, {and} \bibinfo{person}{Xiang-Yang Li}.}
  \bibinfo{year}{2022}\natexlab{}.
\newblock \showarticletitle{Model {Protection}: {Real}-{Time}
  {Privacy}-{Preserving} {Inference} {Service} for {Model} {Privacy} at the
  {Edge}}.
\newblock \bibinfo{journal}{\emph{IEEE Transactions on Dependable and Secure
  Computing}} \bibinfo{volume}{19}, \bibinfo{number}{6} (\bibinfo{date}{Nov.}
  \bibinfo{year}{2022}), \bibinfo{pages}{4270--4284}.
\newblock
\showISSN{1941-0018}
\urldef\tempurl%
\url{https://doi.org/10.1109/TDSC.2021.3126315}
\showDOI{\tempurl}


\bibitem[Hu et~al\mbox{.}(2023)]%
        {hu_secure_2023}
\bibfield{author}{\bibinfo{person}{Bin Hu}, \bibinfo{person}{Yan Wang},
  \bibinfo{person}{Jerry Cheng}, \bibinfo{person}{Tianming Zhao},
  \bibinfo{person}{Yucheng Xie}, \bibinfo{person}{Xiaonan Guo}, {and}
  \bibinfo{person}{Yingying Chen}.} \bibinfo{year}{2023}\natexlab{}.
\newblock \showarticletitle{Secure and {Efficient} {Mobile} {DNN} {Using}
  {Trusted} {Execution} {Environments}}. In
  \bibinfo{booktitle}{\emph{Proceedings of the 2023 {ACM} {Asia} {Conference}
  on {Computer} and {Communications} {Security}}}
  \emph{(\bibinfo{series}{{ASIA} {CCS} '23})}. \bibinfo{publisher}{Association
  for Computing Machinery}, \bibinfo{address}{New York, NY, USA},
  \bibinfo{pages}{274--285}.
\newblock
\showISBNx{979-8-4007-0098-9}
\urldef\tempurl%
\url{https://doi.org/10.1145/3579856.3582820}
\showDOI{\tempurl}


\bibitem[Hua et~al\mbox{.}(2018)]%
        {hua_reverse_2018}
\bibfield{author}{\bibinfo{person}{Weizhe Hua}, \bibinfo{person}{Zhiru Zhang},
  {and} \bibinfo{person}{G.~Edward Suh}.} \bibinfo{year}{2018}\natexlab{}.
\newblock \showarticletitle{Reverse engineering convolutional neural networks
  through side-channel information leaks}. In
  \bibinfo{booktitle}{\emph{Proceedings of the 55th {Annual} {Design}
  {Automation} {Conference}}} \emph{(\bibinfo{series}{{DAC} '18})}.
  \bibinfo{publisher}{Association for Computing Machinery},
  \bibinfo{address}{New York, NY, USA}, \bibinfo{pages}{1--6}.
\newblock
\showISBNx{978-1-4503-5700-5}
\urldef\tempurl%
\url{https://doi.org/10.1145/3195970.3196105}
\showDOI{\tempurl}


\bibitem[Huang et~al\mbox{.}(2025)]%
        {huang_position_2025}
\bibfield{author}{\bibinfo{person}{Hanbo Huang}, \bibinfo{person}{Yihan Li},
  \bibinfo{person}{Bowen Jiang}, \bibinfo{person}{Lin Liu}, \bibinfo{person}{Bo
  Jiang}, \bibinfo{person}{Ruoyu Sun}, \bibinfo{person}{Zhuotao Liu}, {and}
  \bibinfo{person}{Shiyu Liang}.} \bibinfo{year}{2025}\natexlab{}.
\newblock \bibinfo{title}{Position: {On}-{Premises} {LLM} {Deployment}
  {Demands} a {Middle} {Path}: {Preserving} {Privacy} {Without} {Sacrificing}
  {Model} {Confidentiality}}.
\newblock
\newblock
\urldef\tempurl%
\url{https://doi.org/10.48550/arXiv.2410.11182}
\showDOI{\tempurl}


\bibitem[Hunt et~al\mbox{.}(2020)]%
        {hunt_telekine_2020}
\bibfield{author}{\bibinfo{person}{Tyler Hunt}, \bibinfo{person}{Zhipeng Jia},
  \bibinfo{person}{Vance Miller}, \bibinfo{person}{Ariel Szekely},
  \bibinfo{person}{Yige Hu}, \bibinfo{person}{Christopher~J. Rossbach}, {and}
  \bibinfo{person}{Emmett Witchel}.} \bibinfo{year}{2020}\natexlab{}.
\newblock \showarticletitle{Telekine: {Secure} {Computing} with {Cloud}
  {GPUs}}. \bibinfo{pages}{817--833}.
\newblock
\showISBNx{978-1-939133-13-7}
\urldef\tempurl%
\url{https://www.usenix.org/conference/nsdi20/presentation/hunt}
\showURL{%
\tempurl}


\bibitem[Islam et~al\mbox{.}(2023)]%
        {islam_confidential_2023}
\bibfield{author}{\bibinfo{person}{Md~Shihabul Islam}, \bibinfo{person}{Mahmoud
  Zamani}, \bibinfo{person}{Chung~Hwan Kim}, \bibinfo{person}{Latifur Khan},
  {and} \bibinfo{person}{Kevin~W. Hamlen}.} \bibinfo{year}{2023}\natexlab{}.
\newblock \showarticletitle{Confidential {Execution} of {Deep} {Learning}
  {Inference} at the {Untrusted} {Edge} with {ARM} {TrustZone}}. In
  \bibinfo{booktitle}{\emph{Proceedings of the {Thirteenth} {ACM} {Conference}
  on {Data} and {Application} {Security} and {Privacy}}}
  \emph{(\bibinfo{series}{{CODASPY} '23})}. \bibinfo{publisher}{Association for
  Computing Machinery}, \bibinfo{address}{New York, NY, USA},
  \bibinfo{pages}{153--164}.
\newblock
\showISBNx{979-8-4007-0067-5}
\urldef\tempurl%
\url{https://doi.org/10.1145/3577923.3583648}
\showDOI{\tempurl}


\bibitem[Jagielski et~al\mbox{.}(2020)]%
        {jagielski_high_2020}
\bibfield{author}{\bibinfo{person}{Matthew Jagielski},
  \bibinfo{person}{Nicholas Carlini}, \bibinfo{person}{David Berthelot},
  \bibinfo{person}{Alex Kurakin}, {and} \bibinfo{person}{Nicolas Papernot}.}
  \bibinfo{year}{2020}\natexlab{}.
\newblock \showarticletitle{High {Accuracy} and {High} {Fidelity} {Extraction}
  of {Neural} {Networks}}. \bibinfo{pages}{1345--1362}.
\newblock
\showISBNx{978-1-939133-17-5}
\urldef\tempurl%
\url{https://www.usenix.org/conference/usenixsecurity20/presentation/jagielski}
\showURL{%
\tempurl}


\bibitem[Juvekar et~al\mbox{.}(2018)]%
        {juvekar_gazelle_2018}
\bibfield{author}{\bibinfo{person}{Chiraag Juvekar}, \bibinfo{person}{Vinod
  Vaikuntanathan}, {and} \bibinfo{person}{Anantha Chandrakasan}.}
  \bibinfo{year}{2018}\natexlab{}.
\newblock \showarticletitle{\{{GAZELLE}\}: {A} {Low} {Latency} {Framework} for
  {Secure} {Neural} {Network} {Inference}}. \bibinfo{pages}{1651--1669}.
\newblock
\showISBNx{978-1-939133-04-5}
\urldef\tempurl%
\url{https://www.usenix.org/conference/usenixsecurity18/presentation/juvekar}
\showURL{%
\tempurl}


\bibitem[Kaplan et~al\mbox{.}(2021)]%
        {kaplan_memory-encryption-white-paper_nodate}
\bibfield{author}{\bibinfo{person}{David Kaplan}, \bibinfo{person}{Jeremy
  Powell}, {and} \bibinfo{person}{Tom Woller}.}
  \bibinfo{year}{2021}\natexlab{}.
\newblock \bibinfo{booktitle}{\emph{memory-encryption-white-paper}}.
\newblock \bibinfo{type}{{T}echnical {R}eport}.
\newblock
\urldef\tempurl%
\url{https://www.amd.com/content/dam/amd/en/documents/epyc-business-docs/white-papers/memory-encryption-white-paper.pdf}
\showURL{%
\tempurl}


\bibitem[Karumbunathan(2022)]%
        {karumbunathan_nvidia_2022}
\bibfield{author}{\bibinfo{person}{Leela Karumbunathan}.}
  \bibinfo{year}{2022}\natexlab{}.
\newblock \bibinfo{booktitle}{\emph{nvidia jetson agx orin technical brief}}.
\newblock \bibinfo{type}{{T}echnical {R}eport}.
\newblock
\urldef\tempurl%
\url{https://www.nvidia.com/content/dam/en-zz/Solutions/gtcf21/jetson-orin/nvidia-jetson-agx-orin-technical-brief.pdf}
\showURL{%
\tempurl}


\bibitem[Kim et~al\mbox{.}(2020)]%
        {kim_vessels_2020}
\bibfield{author}{\bibinfo{person}{Kyungtae Kim}, \bibinfo{person}{Chung~Hwan
  Kim}, \bibinfo{person}{Junghwan~"John" Rhee}, \bibinfo{person}{Xiao Yu},
  \bibinfo{person}{Haifeng Chen}, \bibinfo{person}{Dave~(Jing) Tian}, {and}
  \bibinfo{person}{Byoungyoung Lee}.} \bibinfo{year}{2020}\natexlab{}.
\newblock \showarticletitle{Vessels: efficient and scalable deep learning
  prediction on trusted processors}. In \bibinfo{booktitle}{\emph{Proceedings
  of the 11th {ACM} {Symposium} on {Cloud} {Computing}}}
  \emph{(\bibinfo{series}{{SoCC} '20})}. \bibinfo{publisher}{Association for
  Computing Machinery}, \bibinfo{address}{New York, NY, USA},
  \bibinfo{pages}{462--476}.
\newblock
\showISBNx{978-1-4503-8137-6}
\urldef\tempurl%
\url{https://doi.org/10.1145/3419111.3421282}
\showDOI{\tempurl}


\bibitem[Kim et~al\mbox{.}(2014)]%
        {kim_flipping_2014}
\bibfield{author}{\bibinfo{person}{Yoongu Kim}, \bibinfo{person}{Ross Daly},
  \bibinfo{person}{Jeremie Kim}, \bibinfo{person}{Chris Fallin},
  \bibinfo{person}{Ji~Hye Lee}, \bibinfo{person}{Donghyuk Lee},
  \bibinfo{person}{Chris Wilkerson}, \bibinfo{person}{Konrad Lai}, {and}
  \bibinfo{person}{Onur Mutlu}.} \bibinfo{year}{2014}\natexlab{}.
\newblock \showarticletitle{Flipping bits in memory without accessing them: an
  experimental study of {DRAM} disturbance errors}.
\newblock \bibinfo{journal}{\emph{SIGARCH Comput. Archit. News}}
  \bibinfo{volume}{42}, \bibinfo{number}{3} (\bibinfo{date}{June}
  \bibinfo{year}{2014}), \bibinfo{pages}{361--372}.
\newblock
\showISSN{0163-5964}
\urldef\tempurl%
\url{https://doi.org/10.1145/2678373.2665726}
\showDOI{\tempurl}


\bibitem[Kocher et~al\mbox{.}(1999)]%
        {kocher_differential_1999}
\bibfield{author}{\bibinfo{person}{Paul Kocher}, \bibinfo{person}{Joshua
  Jaffe}, {and} \bibinfo{person}{Benjamin Jun}.}
  \bibinfo{year}{1999}\natexlab{}.
\newblock \showarticletitle{Differential {Power} {Analysis}}. In
  \bibinfo{booktitle}{\emph{Advances in {Cryptology} — {CRYPTO}’ 99}},
  \bibfield{editor}{\bibinfo{person}{Michael Wiener}} (Ed.).
  \bibinfo{publisher}{Springer}, \bibinfo{address}{Berlin, Heidelberg},
  \bibinfo{pages}{388--397}.
\newblock
\showISBNx{978-3-540-48405-9}
\urldef\tempurl%
\url{https://doi.org/10.1007/3-540-48405-1_25}
\showDOI{\tempurl}


\bibitem[Kocher(1996)]%
        {kocher_timing_1996}
\bibfield{author}{\bibinfo{person}{Paul~C. Kocher}.}
  \bibinfo{year}{1996}\natexlab{}.
\newblock \showarticletitle{Timing {Attacks} on {Implementations} of
  {Diffie}-{Hellman}, {RSA}, {DSS}, and {Other} {Systems}}. In
  \bibinfo{booktitle}{\emph{Advances in {Cryptology} — {CRYPTO} ’96}},
  \bibfield{editor}{\bibinfo{person}{Neal Koblitz}} (Ed.).
  \bibinfo{publisher}{Springer}, \bibinfo{address}{Berlin, Heidelberg},
  \bibinfo{pages}{104--113}.
\newblock
\showISBNx{978-3-540-68697-2}
\urldef\tempurl%
\url{https://doi.org/10.1007/3-540-68697-5_9}
\showDOI{\tempurl}


\bibitem[Kunkel et~al\mbox{.}(2019)]%
        {kunkel_tensorscone_2019}
\bibfield{author}{\bibinfo{person}{Roland Kunkel}, \bibinfo{person}{Do~Le
  Quoc}, \bibinfo{person}{Franz Gregor}, \bibinfo{person}{Sergei Arnautov},
  \bibinfo{person}{Pramod Bhatotia}, {and} \bibinfo{person}{Christof Fetzer}.}
  \bibinfo{year}{2019}\natexlab{}.
\newblock \bibinfo{title}{{TensorSCONE}: {A} {Secure} {TensorFlow} {Framework}
  using {Intel} {SGX}}.
\newblock
\newblock
\urldef\tempurl%
\url{https://doi.org/10.48550/arXiv.1902.04413}
\showDOI{\tempurl}


\bibitem[Lee et~al\mbox{.}(2019)]%
        {lee_occlumency_2019}
\bibfield{author}{\bibinfo{person}{Taegyeong Lee}, \bibinfo{person}{Zhiqi Lin},
  \bibinfo{person}{Saumay Pushp}, \bibinfo{person}{Caihua Li},
  \bibinfo{person}{Yunxin Liu}, \bibinfo{person}{Youngki Lee},
  \bibinfo{person}{Fengyuan Xu}, \bibinfo{person}{Chenren Xu},
  \bibinfo{person}{Lintao Zhang}, {and} \bibinfo{person}{Junehwa Song}.}
  \bibinfo{year}{2019}\natexlab{}.
\newblock \showarticletitle{Occlumency: {Privacy}-preserving {Remote}
  {Deep}-learning {Inference} {Using} {SGX}}. In \bibinfo{booktitle}{\emph{The
  25th {Annual} {International} {Conference} on {Mobile} {Computing} and
  {Networking}}} \emph{(\bibinfo{series}{{MobiCom} '19})}.
  \bibinfo{publisher}{Association for Computing Machinery},
  \bibinfo{address}{New York, NY, USA}, \bibinfo{pages}{1--17}.
\newblock
\showISBNx{978-1-4503-6169-9}
\urldef\tempurl%
\url{https://doi.org/10.1145/3300061.3345447}
\showDOI{\tempurl}


\bibitem[Li et~al\mbox{.}(2024b)]%
        {li_teeslice_2024}
\bibfield{author}{\bibinfo{person}{Ding Li}, \bibinfo{person}{Ziqi Zhang},
  \bibinfo{person}{Mengyu Yao}, \bibinfo{person}{Yifeng Cai},
  \bibinfo{person}{Yao Guo}, {and} \bibinfo{person}{Xiangqun Chen}.}
  \bibinfo{year}{2024}\natexlab{b}.
\newblock \bibinfo{title}{{TEESlice}: {Protecting} {Sensitive} {Neural}
  {Network} {Models} in {Trusted} {Execution} {Environments} {When} {Attackers}
  have {Pre}-{Trained} {Models}}.
\newblock
\newblock
\urldef\tempurl%
\url{https://doi.org/10.48550/arXiv.2411.09945}
\showDOI{\tempurl}


\bibitem[Li et~al\mbox{.}(2023)]%
        {li_secure_2023}
\bibfield{author}{\bibinfo{person}{Fabing Li}, \bibinfo{person}{Xiang Li},
  {and} \bibinfo{person}{Mingyu Gao}.} \bibinfo{year}{2023}\natexlab{}.
\newblock \showarticletitle{Secure {MLaaS} with {Temper}: {Trusted} and
  {Efficient} {Model} {Partitioning} and {Enclave} {Reuse}}. In
  \bibinfo{booktitle}{\emph{Proceedings of the 39th {Annual} {Computer}
  {Security} {Applications} {Conference}}} \emph{(\bibinfo{series}{{ACSAC}
  '23})}. \bibinfo{publisher}{Association for Computing Machinery},
  \bibinfo{address}{New York, NY, USA}, \bibinfo{pages}{621--635}.
\newblock
\showISBNx{979-8-4007-0886-2}
\urldef\tempurl%
\url{https://doi.org/10.1145/3627106.3627145}
\showDOI{\tempurl}


\bibitem[Li et~al\mbox{.}(2024a)]%
        {li_translinkguard_2024}
\bibfield{author}{\bibinfo{person}{Qinfeng Li}, \bibinfo{person}{Zhiqiang
  Shen}, \bibinfo{person}{Zhenghan Qin}, \bibinfo{person}{Yangfan Xie},
  \bibinfo{person}{Xuhong Zhang}, \bibinfo{person}{Tianyu Du}, {and}
  \bibinfo{person}{Jianwei Yin}.} \bibinfo{year}{2024}\natexlab{a}.
\newblock \showarticletitle{{TransLinkGuard}: {Safeguarding} {Transformer}
  {Models} {Against} {Model} {Stealing} in {Edge} {Deployment}}. In
  \bibinfo{booktitle}{\emph{Proceedings of the 32nd {ACM} {International}
  {Conference} on {Multimedia}}}. \bibinfo{pages}{3479--3488}.
\newblock
\urldef\tempurl%
\url{https://doi.org/10.1145/3664647.3680786}
\showDOI{\tempurl}


\bibitem[Liu et~al\mbox{.}(2023)]%
        {liu_mirrornet_2023}
\bibfield{author}{\bibinfo{person}{Ziyu Liu}, \bibinfo{person}{Yukui Luo},
  \bibinfo{person}{Shijin Duan}, \bibinfo{person}{Tong Zhou}, {and}
  \bibinfo{person}{Xiaolin Xu}.} \bibinfo{year}{2023}\natexlab{}.
\newblock \bibinfo{title}{{MirrorNet}: {A} {TEE}-{Friendly} {Framework} for
  {Secure} {On}-device {DNN} {Inference}}.
\newblock
\newblock
\urldef\tempurl%
\url{https://doi.org/10.48550/arXiv.2311.09489}
\showDOI{\tempurl}


\bibitem[Liu et~al\mbox{.}(2024)]%
        {liu_tbnet_2024}
\bibfield{author}{\bibinfo{person}{Ziyu Liu}, \bibinfo{person}{Tong Zhou},
  \bibinfo{person}{Yukui Luo}, {and} \bibinfo{person}{Xiaolin Xu}.}
  \bibinfo{year}{2024}\natexlab{}.
\newblock \bibinfo{title}{{TBNet}: {A} {Neural} {Architectural} {Defense}
  {Framework} {Facilitating} {DNN} {Model} {Protection} in {Trusted}
  {Execution} {Environments}}.
\newblock
\newblock
\urldef\tempurl%
\url{https://doi.org/10.48550/arXiv.2405.03974}
\showDOI{\tempurl}


\bibitem[Mo et~al\mbox{.}(2020)]%
        {mo_darknetz_2020}
\bibfield{author}{\bibinfo{person}{Fan Mo}, \bibinfo{person}{Ali~Shahin
  Shamsabadi}, \bibinfo{person}{Kleomenis Katevas}, \bibinfo{person}{Soteris
  Demetriou}, \bibinfo{person}{Ilias Leontiadis}, \bibinfo{person}{Andrea
  Cavallaro}, {and} \bibinfo{person}{Hamed Haddadi}.}
  \bibinfo{year}{2020}\natexlab{}.
\newblock \showarticletitle{{DarkneTZ}: {Towards} {Model} {Privacy} at the
  {Edge} using {Trusted} {Execution} {Environments}}. In
  \bibinfo{booktitle}{\emph{Proceedings of the 18th {International}
  {Conference} on {Mobile} {Systems}, {Applications}, and {Services}}}.
  \bibinfo{pages}{161--174}.
\newblock
\urldef\tempurl%
\url{https://doi.org/10.1145/3386901.3388946}
\showDOI{\tempurl}


\bibitem[Mo et~al\mbox{.}(2024)]%
        {mo_machine_2024}
\bibfield{author}{\bibinfo{person}{Fan Mo}, \bibinfo{person}{Zahra Tarkhani},
  {and} \bibinfo{person}{Hamed Haddadi}.} \bibinfo{year}{2024}\natexlab{}.
\newblock \showarticletitle{Machine {Learning} with {Confidential} {Computing}:
  {A} {Systematization} of {Knowledge}}.
\newblock \bibinfo{journal}{\emph{ACM Comput. Surv.}} \bibinfo{volume}{56},
  \bibinfo{number}{11} (\bibinfo{date}{June} \bibinfo{year}{2024}),
  \bibinfo{pages}{281:1--281:40}.
\newblock
\showISSN{0360-0300}
\urldef\tempurl%
\url{https://doi.org/10.1145/3670007}
\showDOI{\tempurl}


\bibitem[Narra et~al\mbox{.}(2019)]%
        {narra_privacy-preserving_2019}
\bibfield{author}{\bibinfo{person}{Krishna~Giri Narra},
  \bibinfo{person}{Zhifeng Lin}, \bibinfo{person}{Yongqin Wang},
  \bibinfo{person}{Keshav Balasubramaniam}, {and} \bibinfo{person}{Murali
  Annavaram}.} \bibinfo{year}{2019}\natexlab{}.
\newblock \bibinfo{title}{Privacy-{Preserving} {Inference} in {Machine}
  {Learning} {Services} {Using} {Trusted} {Execution} {Environments}}.
\newblock
\newblock
\urldef\tempurl%
\url{https://doi.org/10.48550/arXiv.1912.03485}
\showDOI{\tempurl}


\bibitem[Nayan et~al\mbox{.}(2024)]%
        {nayan_sok_nodate}
\bibfield{author}{\bibinfo{person}{Tushar Nayan}, \bibinfo{person}{Qiming Guo},
  {and} \bibinfo{person}{Mohammed~Al Duniawi}.}
  \bibinfo{year}{2024}\natexlab{}.
\newblock \showarticletitle{{SoK}: {All} {You} {Need} to {Know} {About}
  {On}-{Device} {ML} {Model} {Extraction} - {The} {Gap} {Between} {Research}
  and {Practice}}.
\newblock  (\bibinfo{year}{2024}).
\newblock


\bibitem[Ohrimenko et~al\mbox{.}(2016)]%
        {ohrimenko_oblivious_2016}
\bibfield{author}{\bibinfo{person}{Olga Ohrimenko}, \bibinfo{person}{Felix
  Schuster}, \bibinfo{person}{Cédric Fournet}, \bibinfo{person}{Aastha Mehta},
  \bibinfo{person}{Sebastian Nowozin}, \bibinfo{person}{Kapil Vaswani}, {and}
  \bibinfo{person}{Manuel Costa}.} \bibinfo{year}{2016}\natexlab{}.
\newblock \showarticletitle{Oblivious \{{Multi}-{Party}\} {Machine} {Learning}
  on {Trusted} {Processors}}. \bibinfo{pages}{619--636}.
\newblock
\showISBNx{978-1-931971-32-4}
\urldef\tempurl%
\url{https://www.usenix.org/conference/usenixsecurity16/technical-sessions/presentation/ohrimenko}
\showURL{%
\tempurl}


\bibitem[Oliynyk et~al\mbox{.}(2025)]%
        {oliynyk_i_2025}
\bibfield{author}{\bibinfo{person}{Daryna Oliynyk}, \bibinfo{person}{Rudolf
  Mayer}, \bibinfo{person}{Kathrin Grosse}, {and} \bibinfo{person}{Andreas
  Rauber}.} \bibinfo{year}{2025}\natexlab{}.
\newblock \bibinfo{title}{I {Stolenly} {Swear} {That} {I} {Am} {Up} to ({No})
  {Good}: {Design} and {Evaluation} of {Model} {Stealing} {Attacks}}.
\newblock
\newblock
\urldef\tempurl%
\url{https://doi.org/10.48550/arXiv.2508.21654}
\showDOI{\tempurl}


\bibitem[OpenAI(2025a)]%
        {openai_chatgpt_nodate}
\bibfield{author}{\bibinfo{person}{OpenAI}.} \bibinfo{year}{2025}\natexlab{a}.
\newblock \bibinfo{title}{{ChatGPT}}.
\newblock
\newblock
\urldef\tempurl%
\url{https://chatgpt.com/?locale=en-US}
\showURL{%
\tempurl}


\bibitem[OpenAI(2025b)]%
        {openai_chatgptforenterprise_2025}
\bibfield{author}{\bibinfo{person}{OpenAI}.} \bibinfo{year}{2025}\natexlab{b}.
\newblock \bibinfo{title}{{ChatGPT for Enterprise}}.
\newblock
\newblock
\urldef\tempurl%
\url{https://chatgpt.com/business/enterprise/}
\showURL{%
\tempurl}


\bibitem[Quoc et~al\mbox{.}(2020)]%
        {quoc_securetf_2020}
\bibfield{author}{\bibinfo{person}{Do~Le Quoc}, \bibinfo{person}{Franz Gregor},
  \bibinfo{person}{Sergei Arnautov}, \bibinfo{person}{Roland Kunkel},
  \bibinfo{person}{Pramod Bhatotia}, {and} \bibinfo{person}{Christof Fetzer}.}
  \bibinfo{year}{2020}\natexlab{}.
\newblock \showarticletitle{{secureTF}: {A} {Secure} {TensorFlow} {Framework}}.
  In \bibinfo{booktitle}{\emph{Proceedings of the 21st {International}
  {Middleware} {Conference}}} \emph{(\bibinfo{series}{Middleware '20})}.
  \bibinfo{publisher}{Association for Computing Machinery},
  \bibinfo{address}{New York, NY, USA}, \bibinfo{pages}{44--59}.
\newblock
\showISBNx{978-1-4503-8153-6}
\urldef\tempurl%
\url{https://doi.org/10.1145/3423211.3425687}
\showDOI{\tempurl}


\bibitem[Rakin et~al\mbox{.}(2022)]%
        {rakin_deepsteal_2022}
\bibfield{author}{\bibinfo{person}{Adnan~Siraj Rakin},
  \bibinfo{person}{Md~Hafizul~Islam Chowdhuryy}, \bibinfo{person}{Fan Yao},
  {and} \bibinfo{person}{Deliang Fan}.} \bibinfo{year}{2022}\natexlab{}.
\newblock \showarticletitle{{DeepSteal}: {Advanced} {Model} {Extractions}
  {Leveraging} {Efficient} {Weight} {Stealing} in {Memories}}. In
  \bibinfo{booktitle}{\emph{2022 {IEEE} {Symposium} on {Security} and {Privacy}
  ({SP})}}. \bibinfo{pages}{1157--1174}.
\newblock
\urldef\tempurl%
\url{https://doi.org/10.1109/SP46214.2022.9833743}
\showDOI{\tempurl}


\bibitem[Rakin et~al\mbox{.}(2021)]%
        {rakin_t-bfa_2021}
\bibfield{author}{\bibinfo{person}{Adnan~Siraj Rakin}, \bibinfo{person}{Zhezhi
  He}, \bibinfo{person}{Jingtao Li}, \bibinfo{person}{Fan Yao},
  \bibinfo{person}{Chaitali Chakrabarti}, {and} \bibinfo{person}{Deliang Fan}.}
  \bibinfo{year}{2021}\natexlab{}.
\newblock \bibinfo{title}{T-{BFA}: {Targeted} {Bit}-{Flip} {Adversarial}
  {Weight} {Attack}}.
\newblock
\newblock
\urldef\tempurl%
\url{https://doi.org/10.48550/arXiv.2007.12336}
\showDOI{\tempurl}


\bibitem[Refael et~al\mbox{.}(2024)]%
        {refael_slip_2024}
\bibfield{author}{\bibinfo{person}{Yehonathan Refael}, \bibinfo{person}{Adam
  Hakim}, \bibinfo{person}{Lev Greenberg}, \bibinfo{person}{Tal Aviv},
  \bibinfo{person}{Satya Lokam}, \bibinfo{person}{Ben Fishman}, {and}
  \bibinfo{person}{Shachar Seidman}.} \bibinfo{year}{2024}\natexlab{}.
\newblock \bibinfo{title}{{SLIP}: {Securing} {LLMs} {IP} {Using} {Weights}
  {Decomposition}}.
\newblock
\newblock
\urldef\tempurl%
\url{https://doi.org/10.48550/arXiv.2407.10886}
\showDOI{\tempurl}


\bibitem[Rolnick and Kording(2019)]%
        {rolnick_reverse-engineering_2019}
\bibfield{author}{\bibinfo{person}{David Rolnick} {and}
  \bibinfo{person}{Konrad~P. Kording}.} \bibinfo{year}{2019}\natexlab{}.
\newblock \bibinfo{title}{Reverse-{Engineering} {Deep} {ReLU} {Networks}}.
\newblock
\newblock
\urldef\tempurl%
\url{https://arxiv.org/abs/1910.00744v2}
\showURL{%
\tempurl}


\bibitem[Sanyal et~al\mbox{.}(2022)]%
        {sanyal_towards_2022}
\bibfield{author}{\bibinfo{person}{Sunandini Sanyal}, \bibinfo{person}{Sravanti
  Addepalli}, {and} \bibinfo{person}{R.~Venkatesh Babu}.}
  \bibinfo{year}{2022}\natexlab{}.
\newblock \bibinfo{title}{Towards {Data}-{Free} {Model} {Stealing} in a {Hard}
  {Label} {Setting}}.
\newblock
\newblock
\urldef\tempurl%
\url{https://doi.org/10.48550/arXiv.2204.11022}
\showDOI{\tempurl}


\bibitem[Schlögl and Böhme(2020)]%
        {schlogl_ennclave_2020}
\bibfield{author}{\bibinfo{person}{Alexander Schlögl} {and}
  \bibinfo{person}{Rainer Böhme}.} \bibinfo{year}{2020}\natexlab{}.
\newblock \showarticletitle{{eNNclave}: {Offline} {Inference} with {Model}
  {Confidentiality}}. In \bibinfo{booktitle}{\emph{Proceedings of the 13th
  {ACM} {Workshop} on {Artificial} {Intelligence} and {Security}}}
  \emph{(\bibinfo{series}{{AISec}'20})}. \bibinfo{publisher}{Association for
  Computing Machinery}, \bibinfo{address}{New York, NY, USA},
  \bibinfo{pages}{93--104}.
\newblock
\showISBNx{978-1-4503-8094-2}
\urldef\tempurl%
\url{https://doi.org/10.1145/3411508.3421376}
\showDOI{\tempurl}


\bibitem[Schneier and Mudge(1998)]%
        {schneier_cryptanalysis_1998}
\bibfield{author}{\bibinfo{person}{Bruce Schneier} {and}
  \bibinfo{person}{Mudge}.} \bibinfo{year}{1998}\natexlab{}.
\newblock \showarticletitle{Cryptanalysis of {Microsoft}'s point-to-point
  tunneling protocol ({PPTP})}. In \bibinfo{booktitle}{\emph{Proceedings of the
  5th {ACM} conference on {Computer} and communications security}}
  \emph{(\bibinfo{series}{{CCS} '98})}. \bibinfo{publisher}{Association for
  Computing Machinery}, \bibinfo{address}{New York, NY, USA},
  \bibinfo{pages}{132--141}.
\newblock
\showISBNx{978-1-58113-007-2}
\urldef\tempurl%
\url{https://doi.org/10.1145/288090.288119}
\showDOI{\tempurl}


\bibitem[SentinelOne(2025)]%
        {sentinelone_chatgpt_2025}
\bibfield{author}{\bibinfo{person}{SentinelOne}.}
  \bibinfo{year}{2025}\natexlab{}.
\newblock \bibinfo{title}{{ChatGPT} {Security} {Risks}: {All} {You} {Need} to
  {Know}}.
\newblock
\newblock
\urldef\tempurl%
\url{https://www.sentinelone.com/cybersecurity-101/data-and-ai/chatgpt-security-risks/}
\showURL{%
\tempurl}


\bibitem[Shen et~al\mbox{.}(2022)]%
        {shen_soter_2022}
\bibfield{author}{\bibinfo{person}{Tianxiang Shen}, \bibinfo{person}{Ji Qi},
  \bibinfo{person}{Jianyu Jiang}, \bibinfo{person}{Xian Wang},
  \bibinfo{person}{Siyuan Wen}, \bibinfo{person}{Xusheng Chen},
  \bibinfo{person}{Shixiong Zhao}, {and} \bibinfo{person}{Xiapu Luo}.}
  \bibinfo{year}{2022}\natexlab{}.
\newblock \showarticletitle{Soter: {Guarding} {Black}-box {Inference} for
  {General} {Neural} {Networks} at the {Edge}}.
\newblock  (\bibinfo{year}{2022}).
\newblock


\bibitem[Siby et~al\mbox{.}(2024)]%
        {siby_guarantee_2024}
\bibfield{author}{\bibinfo{person}{Sandra Siby}, \bibinfo{person}{Sina
  Abdollahi}, \bibinfo{person}{Mohammad Maheri}, \bibinfo{person}{Marios
  Kogias}, {and} \bibinfo{person}{Hamed Haddadi}.}
  \bibinfo{year}{2024}\natexlab{}.
\newblock \bibinfo{title}{{GuaranTEE}: {Towards} {Attestable} and {Private}
  {ML} with {CCA}}.
\newblock
\newblock
\urldef\tempurl%
\url{https://doi.org/10.48550/arXiv.2404.00190}
\showDOI{\tempurl}


\bibitem[Sorensen and Khlaaf(2024)]%
        {sorensen_leftoverlocals_2024}
\bibfield{author}{\bibinfo{person}{Tyler Sorensen} {and} \bibinfo{person}{Heidy
  Khlaaf}.} \bibinfo{year}{2024}\natexlab{}.
\newblock \bibinfo{title}{{LeftoverLocals}: {Listening} to {LLM} {Responses}
  {Through} {Leaked} {GPU} {Local} {Memory}}.
\newblock
\newblock
\urldef\tempurl%
\url{https://doi.org/10.48550/arXiv.2401.16603}
\showDOI{\tempurl}


\bibitem[Stinson and Paterson(2018)]%
        {stinson_cryptography_2018}
\bibfield{author}{\bibinfo{person}{Douglas~Robert Stinson} {and}
  \bibinfo{person}{Maura Paterson}.} \bibinfo{year}{2018}\natexlab{}.
\newblock \bibinfo{booktitle}{\emph{Cryptography: {Theory} and {Practice}}
  (\bibinfo{edition}{4} ed.)}.
\newblock \bibinfo{publisher}{Chapman and Hall/CRC}, \bibinfo{address}{New
  York}.
\newblock
\showISBNx{978-1-315-28249-7}


\bibitem[Stoica et~al\mbox{.}(2017)]%
        {stoica_berkeley_2017}
\bibfield{author}{\bibinfo{person}{Ion Stoica}, \bibinfo{person}{Dawn Song},
  \bibinfo{person}{Raluca~Ada Popa}, \bibinfo{person}{David~A. Patterson},
  \bibinfo{person}{Michael~W. Mahoney}, \bibinfo{person}{Randy~H. Katz},
  \bibinfo{person}{Anthony~D. Joseph}, \bibinfo{person}{Michael Jordan},
  \bibinfo{person}{Joseph~M. Hellerstein}, \bibinfo{person}{Joseph Gonzalez},
  \bibinfo{person}{Ken Goldberg}, \bibinfo{person}{Ali Ghodsi},
  \bibinfo{person}{David~E. Culler}, {and} \bibinfo{person}{Pieter Abbeel}.}
  \bibinfo{year}{2017}\natexlab{}.
\newblock \bibinfo{booktitle}{\emph{A {Berkeley} {View} of {Systems}
  {Challenges} for {AI}}}.
\newblock \bibinfo{type}{{T}echnical {R}eport} UCB/EECS-2017-159.
\newblock
\urldef\tempurl%
\url{http://www2.eecs.berkeley.edu/Pubs/TechRpts/2017/EECS-2017-159.html}
\showURL{%
\tempurl}


\bibitem[Sun et~al\mbox{.}(2025a)]%
        {sun_tensorshield_2025}
\bibfield{author}{\bibinfo{person}{Tong Sun}, \bibinfo{person}{Bowen Jiang},
  \bibinfo{person}{Hailong Lin}, \bibinfo{person}{Borui Li},
  \bibinfo{person}{Yixiao Teng}, \bibinfo{person}{Yi Gao}, {and}
  \bibinfo{person}{Wei Dong}.} \bibinfo{year}{2025}\natexlab{a}.
\newblock \bibinfo{title}{{TensorShield}: {Safeguarding} {On}-{Device}
  {Inference} by {Shielding} {Critical} {DNN} {Tensors} with {TEE}}.
\newblock
\newblock
\urldef\tempurl%
\url{https://doi.org/10.48550/arXiv.2505.22735}
\showDOI{\tempurl}


\bibitem[Sun et~al\mbox{.}(2025b)]%
        {sun_tsqp_2025}
\bibfield{author}{\bibinfo{person}{Yu Sun}, \bibinfo{person}{Gaojian Xiong},
  \bibinfo{person}{Jianhua Liu}, \bibinfo{person}{Zheng Liu}, {and}
  \bibinfo{person}{Jian Cui}.} \bibinfo{year}{2025}\natexlab{b}.
\newblock \showarticletitle{{TSQP}: {Safeguarding} {Real}-{Time} {Inference}
  for {Quantization} {Neural} {Networks} on {Edge} {Devices}}. In
  \bibinfo{booktitle}{\emph{2025 {IEEE} {Symposium} on {Security} and {Privacy}
  ({SP})}}. \bibinfo{publisher}{IEEE Computer Society}, \bibinfo{address}{Los
  Alamitos, CA, USA}, \bibinfo{pages}{2114--2132}.
\newblock
\urldef\tempurl%
\url{https://doi.org/10.1109/SP61157.2025.00001}
\showDOI{\tempurl}


\bibitem[Sun et~al\mbox{.}(2023)]%
        {sun_shadownet_2023}
\bibfield{author}{\bibinfo{person}{Zhichuang Sun}, \bibinfo{person}{Ruimin
  Sun}, \bibinfo{person}{Changming Liu}, \bibinfo{person}{Amrita~Roy
  Chowdhury}, \bibinfo{person}{Long Lu}, {and} \bibinfo{person}{Somesh Jha}.}
  \bibinfo{year}{2023}\natexlab{}.
\newblock \bibinfo{title}{{ShadowNet}: {A} {Secure} and {Efficient} {On}-device
  {Model} {Inference} {System} for {Convolutional} {Neural} {Networks}}.
\newblock
\newblock
\urldef\tempurl%
\url{https://doi.org/10.48550/arXiv.2011.05905}
\showDOI{\tempurl}


\bibitem[Sun et~al\mbox{.}(2021)]%
        {sun_mind_2021}
\bibfield{author}{\bibinfo{person}{Zhichuang Sun}, \bibinfo{person}{Ruimin
  Sun}, \bibinfo{person}{Long Lu}, {and} \bibinfo{person}{Alan Mislove}.}
  \bibinfo{year}{2021}\natexlab{}.
\newblock \bibinfo{title}{Mind {Your} {Weight}(s): {A} {Large}-scale {Study} on
  {Insufficient} {Machine} {Learning} {Model} {Protection} in {Mobile} {Apps}}.
\newblock
\newblock
\urldef\tempurl%
\url{https://doi.org/10.48550/arXiv.2002.07687}
\showDOI{\tempurl}


\bibitem[Tesla(2025)]%
        {tesla_autopilot_nodate}
\bibfield{author}{\bibinfo{person}{Tesla}.} \bibinfo{year}{2025}\natexlab{}.
\newblock \bibinfo{title}{Autopilot {\textbar} {Tesla} {Support}}.
\newblock
\newblock
\urldef\tempurl%
\url{https://www.tesla.com/support/autopilot}
\showURL{%
\tempurl}


\bibitem[Tramèr and Boneh(2019)]%
        {tramer_slalom_2019}
\bibfield{author}{\bibinfo{person}{Florian Tramèr} {and} \bibinfo{person}{Dan
  Boneh}.} \bibinfo{year}{2019}\natexlab{}.
\newblock \bibinfo{title}{Slalom: {Fast}, {Verifiable} and {Private}
  {Execution} of {Neural} {Networks} in {Trusted} {Hardware}}.
\newblock
\newblock
\urldef\tempurl%
\url{https://doi.org/10.48550/arXiv.1806.03287}
\showDOI{\tempurl}


\bibitem[Tramèr et~al\mbox{.}(2016)]%
        {tramer_stealing_2016}
\bibfield{author}{\bibinfo{person}{Florian Tramèr}, \bibinfo{person}{Fan
  Zhang}, \bibinfo{person}{Ari Juels}, \bibinfo{person}{Michael~K. Reiter},
  {and} \bibinfo{person}{Thomas Ristenpart}.} \bibinfo{year}{2016}\natexlab{}.
\newblock \bibinfo{title}{Stealing {Machine} {Learning} {Models} via
  {Prediction} {APIs}}.
\newblock
\newblock
\urldef\tempurl%
\url{https://doi.org/10.48550/arXiv.1609.02943}
\showDOI{\tempurl}


\bibitem[Vanhoef and Piessens(2017)]%
        {vanhoef_key_2017}
\bibfield{author}{\bibinfo{person}{Mathy Vanhoef} {and} \bibinfo{person}{Frank
  Piessens}.} \bibinfo{year}{2017}\natexlab{}.
\newblock \showarticletitle{Key {Reinstallation} {Attacks}: {Forcing} {Nonce}
  {Reuse} in {WPA2}}. In \bibinfo{booktitle}{\emph{Proceedings of the 2017
  {ACM} {SIGSAC} {Conference} on {Computer} and {Communications} {Security}}}
  \emph{(\bibinfo{series}{{CCS} '17})}. \bibinfo{publisher}{Association for
  Computing Machinery}, \bibinfo{address}{New York, NY, USA},
  \bibinfo{pages}{1313--1328}.
\newblock
\showISBNx{978-1-4503-4946-8}
\urldef\tempurl%
\url{https://doi.org/10.1145/3133956.3134027}
\showDOI{\tempurl}


\bibitem[VanNostrand et~al\mbox{.}(2019)]%
        {vannostrand_confidential_2019}
\bibfield{author}{\bibinfo{person}{Peter~M. VanNostrand},
  \bibinfo{person}{Ioannis Kyriazis}, \bibinfo{person}{Michelle Cheng},
  \bibinfo{person}{Tian Guo}, {and} \bibinfo{person}{Robert~J. Walls}.}
  \bibinfo{year}{2019}\natexlab{}.
\newblock \bibinfo{title}{Confidential {Deep} {Learning}: {Executing}
  {Proprietary} {Models} on {Untrusted} {Devices}}.
\newblock
\newblock
\urldef\tempurl%
\url{https://doi.org/10.48550/arXiv.1908.10730}
\showDOI{\tempurl}


\bibitem[Vaswani et~al\mbox{.}(2023)]%
        {vaswani_confidential_2023}
\bibfield{author}{\bibinfo{person}{Kapil Vaswani}, \bibinfo{person}{Stavros
  Volos}, \bibinfo{person}{Cédric Fournet}, \bibinfo{person}{Antonio~Nino
  Diaz}, \bibinfo{person}{Ken Gordon}, \bibinfo{person}{Balaji Vembu},
  \bibinfo{person}{Sam Webster}, \bibinfo{person}{David Chisnall},
  \bibinfo{person}{Saurabh Kulkarni}, \bibinfo{person}{Graham Cunningham},
  \bibinfo{person}{Richard Osborne}, {and} \bibinfo{person}{Daniel Wilkinson}.}
  \bibinfo{year}{2023}\natexlab{}.
\newblock \showarticletitle{Confidential {Computing} within an {AI}
  {Accelerator}}. \bibinfo{pages}{501--518}.
\newblock
\showISBNx{978-1-939133-35-9}
\urldef\tempurl%
\url{https://www.usenix.org/conference/atc23/presentation/vaswani}
\showURL{%
\tempurl}


\bibitem[Volos et~al\mbox{.}(2018)]%
        {volos_graviton_2018}
\bibfield{author}{\bibinfo{person}{Stavros Volos}, \bibinfo{person}{Kapil
  Vaswani}, {and} \bibinfo{person}{Rodrigo Bruno}.}
  \bibinfo{year}{2018}\natexlab{}.
\newblock \showarticletitle{Graviton: {Trusted} {Execution} {Environments} on
  {GPUs}}. \bibinfo{pages}{681--696}.
\newblock
\showISBNx{978-1-939133-08-3}
\urldef\tempurl%
\url{https://www.usenix.org/conference/osdi18/presentation/volos}
\showURL{%
\tempurl}


\bibitem[Wang(2025)]%
        {wang_game_nodate}
\bibfield{author}{\bibinfo{person}{Pengli Wang}.}
  \bibinfo{year}{2025}\natexlab{}.
\newblock \showarticletitle{Game of {Arrows}: {On} the ({In}-){Security} of
  {Weight} {Obfuscation} for {On}-{Device} {TEE}-{Shielded} {LLM} {Partition}
  {Algorithms}}.
\newblock  (\bibinfo{year}{2025}).
\newblock


\bibitem[Wei et~al\mbox{.}(2020)]%
        {wei_leaky_2020}
\bibfield{author}{\bibinfo{person}{Junyi Wei}, \bibinfo{person}{Yicheng Zhang},
  \bibinfo{person}{Zhe Zhou}, \bibinfo{person}{Zhou Li}, {and}
  \bibinfo{person}{Mohammad~Abdullah Al~Faruque}.}
  \bibinfo{year}{2020}\natexlab{}.
\newblock \showarticletitle{Leaky {DNN}: {Stealing} {Deep}-{Learning} {Model}
  {Secret} with {GPU} {Context}-{Switching} {Side}-{Channel}}. In
  \bibinfo{booktitle}{\emph{2020 50th {Annual} {IEEE}/{IFIP} {International}
  {Conference} on {Dependable} {Systems} and {Networks} ({DSN})}}.
  \bibinfo{pages}{125--137}.
\newblock
\urldef\tempurl%
\url{https://doi.org/10.1109/DSN48063.2020.00031}
\showDOI{\tempurl}


\bibitem[Wilhelm(2023)]%
        {wilhelm_amazon_2023}
\bibfield{author}{\bibinfo{person}{Henry Wilhelm}.}
  \bibinfo{year}{2023}\natexlab{}.
\newblock \bibinfo{title}{Amazon unveils a new lineup of {Echo}
  devices—here’s your first look}.
\newblock
\newblock
\urldef\tempurl%
\url{https://www.aboutamazon.com/news/devices/new-amazon-echo-devices-echo-buds-echo-pop}
\showURL{%
\tempurl}


\bibitem[Xu and Fang(2024)]%
        {xu_tempo_2024}
\bibfield{author}{\bibinfo{person}{Rongwu Xu} {and} \bibinfo{person}{Zhixuan
  Fang}.} \bibinfo{year}{2024}\natexlab{}.
\newblock \bibinfo{title}{Tempo: {Confidentiality} {Preservation} in
  {Cloud}-{Based} {Neural} {Network} {Training}}.
\newblock
\newblock
\urldef\tempurl%
\url{https://doi.org/10.48550/arXiv.2401.11531}
\showDOI{\tempurl}


\bibitem[Yan et~al\mbox{.}(2020)]%
        {yan_cache_2020}
\bibfield{author}{\bibinfo{person}{Mengjia Yan},
  \bibinfo{person}{Christopher~W. Fletcher}, {and} \bibinfo{person}{Josep
  Torrellas}.} \bibinfo{year}{2020}\natexlab{}.
\newblock \showarticletitle{Cache {Telepathy}: {Leveraging} {Shared} {Resource}
  {Attacks} to {Learn} {DNN} {Architectures}}. \bibinfo{pages}{2003--2020}.
\newblock
\showISBNx{978-1-939133-17-5}
\urldef\tempurl%
\url{https://www.usenix.org/conference/usenixsecurity20/presentation/yan}
\showURL{%
\tempurl}


\bibitem[Yang et~al\mbox{.}(2024)]%
        {yang_first_2024}
\bibfield{author}{\bibinfo{person}{Huan Yang}, \bibinfo{person}{Deyu Zhang},
  \bibinfo{person}{Yudong Zhao}, \bibinfo{person}{Yuanchun Li}, {and}
  \bibinfo{person}{Yunxin Liu}.} \bibinfo{year}{2024}\natexlab{}.
\newblock \bibinfo{title}{A {First} {Look} {At} {Efficient} {And} {Secure}
  {On}-{Device} {LLM} {Inference} {Against} {KV} {Leakage}}.
\newblock
\newblock
\urldef\tempurl%
\url{https://doi.org/10.48550/arXiv.2409.04040}
\showDOI{\tempurl}


\bibitem[Yarom and Falkner(2014)]%
        {yarom_flushreload_2014}
\bibfield{author}{\bibinfo{person}{Yuval Yarom} {and} \bibinfo{person}{Katrina
  Falkner}.} \bibinfo{year}{2014}\natexlab{}.
\newblock \showarticletitle{{FLUSH}+{RELOAD}: {A} {High} {Resolution}, {Low}
  {Noise}, {L3} {Cache} {Side}-{Channel} {Attack}}. \bibinfo{pages}{719--732}.
\newblock
\showISBNx{978-1-931971-15-7}
\urldef\tempurl%
\url{https://www.usenix.org/conference/usenixsecurity14/technical-sessions/presentation/yarom}
\showURL{%
\tempurl}


\bibitem[Yuan et~al\mbox{.}(2024b)]%
        {yuan_secure_2024}
\bibfield{author}{\bibinfo{person}{Mu Yuan}, \bibinfo{person}{Lan Zhang}, {and}
  \bibinfo{person}{Xiang-Yang Li}.} \bibinfo{year}{2024}\natexlab{b}.
\newblock \bibinfo{title}{Secure {Transformer} {Inference} {Protocol}}.
\newblock
\newblock
\urldef\tempurl%
\url{https://doi.org/10.48550/arXiv.2312.00025}
\showDOI{\tempurl}


\bibitem[Yuan et~al\mbox{.}(2024a)]%
        {yuan_ciphersteal_2024}
\bibfield{author}{\bibinfo{person}{Yuanyuan Yuan}, \bibinfo{person}{Zhibo Liu},
  \bibinfo{person}{Sen Deng}, \bibinfo{person}{Yanzuo Chen},
  \bibinfo{person}{Shuai Wang}, \bibinfo{person}{Yinqian Zhang}, {and}
  \bibinfo{person}{Zhendong Su}.} \bibinfo{year}{2024}\natexlab{a}.
\newblock \showarticletitle{{CipherSteal}: {Stealing} {Input} {Data} from
  {TEE}-{Shielded} {Neural} {Networks} with {Ciphertext} {Side} {Channels}}.
  \bibinfo{publisher}{IEEE Computer Society}, \bibinfo{pages}{4136--4154}.
\newblock
\showISBNx{979-8-3315-2236-0}
\urldef\tempurl%
\url{https://doi.org/10.1109/SP61157.2025.00079}
\showDOI{\tempurl}


\bibitem[Zhang et~al\mbox{.}(2023)]%
        {zhang_no_2023}
\bibfield{author}{\bibinfo{person}{Ziqi Zhang}, \bibinfo{person}{Chen Gong},
  \bibinfo{person}{Yifeng Cai}, \bibinfo{person}{Yuanyuan Yuan},
  \bibinfo{person}{Bingyan Liu}, \bibinfo{person}{Ding Li},
  \bibinfo{person}{Yao Guo}, {and} \bibinfo{person}{Xiangqun Chen}.}
  \bibinfo{year}{2023}\natexlab{}.
\newblock \bibinfo{title}{No {Privacy} {Left} {Outside}: {On} the
  ({In}-){Security} of {TEE}-{Shielded} {DNN} {Partition} for {On}-{Device}
  {ML}}.
\newblock
\newblock
\urldef\tempurl%
\url{https://doi.org/10.48550/arXiv.2310.07152}
\showDOI{\tempurl}


\bibitem[Zhang et~al\mbox{.}(2024)]%
        {zhang_groupcover_2024}
\bibfield{author}{\bibinfo{person}{Zheng Zhang}, \bibinfo{person}{Na Wang},
  \bibinfo{person}{Ziqi Zhang}, \bibinfo{person}{Yao Zhang},
  \bibinfo{person}{Tianyi Zhang}, \bibinfo{person}{Jianwei Liu}, {and}
  \bibinfo{person}{Ye Wu}.} \bibinfo{year}{2024}\natexlab{}.
\newblock \showarticletitle{{GroupCover}: {A} {Secure}, {Efficient} and
  {Scalable} {Inference} {Framework} for {On}-device {Model} {Protection} based
  on {TEEs}}. In \bibinfo{booktitle}{\emph{Proceedings of the 41st
  {International} {Conference} on {Machine} {Learning}}}.
  \bibinfo{publisher}{PMLR}, \bibinfo{pages}{59992--60003}.
\newblock
\urldef\tempurl%
\url{https://proceedings.mlr.press/v235/zhang24bn.html}
\showURL{%
\tempurl}


\bibitem[Zhao et~al\mbox{.}(2019)]%
        {zhao_fault_2019}
\bibfield{author}{\bibinfo{person}{Pu Zhao}, \bibinfo{person}{Siyue Wang},
  \bibinfo{person}{Cheng Gongye}, \bibinfo{person}{Yanzhi Wang},
  \bibinfo{person}{Yunsi Fei}, {and} \bibinfo{person}{Xue Lin}.}
  \bibinfo{year}{2019}\natexlab{}.
\newblock \showarticletitle{Fault {Sneaking} {Attack}: a {Stealthy} {Framework}
  for {Misleading} {Deep} {Neural} {Networks}}. In
  \bibinfo{booktitle}{\emph{Proceedings of the 56th {Annual} {Design}
  {Automation} {Conference} 2019}} \emph{(\bibinfo{series}{{DAC} '19})}.
  \bibinfo{publisher}{Association for Computing Machinery},
  \bibinfo{address}{New York, NY, USA}, \bibinfo{pages}{1--6}.
\newblock
\showISBNx{978-1-4503-6725-7}
\urldef\tempurl%
\url{https://doi.org/10.1145/3316781.3317825}
\showDOI{\tempurl}


\bibitem[Zhu et~al\mbox{.}(2024)]%
        {zhu_confidential_2024}
\bibfield{author}{\bibinfo{person}{Jianwei Zhu}, \bibinfo{person}{Hang Yin},
  \bibinfo{person}{Peng Deng}, \bibinfo{person}{Aline Almeida}, {and}
  \bibinfo{person}{Shunfan Zhou}.} \bibinfo{year}{2024}\natexlab{}.
\newblock \bibinfo{title}{Confidential {Computing} on {NVIDIA} {Hopper} {GPUs}:
  {A} {Performance} {Benchmark} {Study}}.
\newblock
\newblock
\urldef\tempurl%
\url{https://doi.org/10.48550/arXiv.2409.03992}
\showDOI{\tempurl}


\end{thebibliography}

\clearpage
\appendix

\section{Appendix: Derivation of the Rank Deficiency Probability}
\label{app:derivation_rank_def}

We want to find the probability that a randomly generated $K \times K$ matrix $M$, whose entries are chosen uniformly from a finite field $\mathbb{F}_P$ with $P$ elements (i.e., each entry of matrix $M$ can be any integer from $0$ to $P-1$ with equal probability), is rank deficient (i.e., singular or has columns that are linearly dependent).

It is easier to first calculate the probability of the complementary event: that the matrix $M$ has \textbf{full rank} (i.e., is invertible). We can then find the desired probability using the relation:
\begin{equation*}
P(\text{rank deficient}) = 1 - P(\text{full rank})
\end{equation*}

Using the chain rule of probability, we can express this probability as:
\begin{align*}
    P(\text{full rank}) &= P(K \text{ cols are independent}) \\
    &= P(\text{col } K \text{ is indep. } | \text{ first } K-1 \text{ are indep.}) \\
    & \quad \times P(\text{first } K-1 \text{ cols are indep.}) \\
    &= P(\text{col } K \text{ is indep. } | \text{ first } K-1 \text{ are indep.}) \\
    & \quad \times P(\text{col } K-1 \text{ is indep. } | \text{ first } K-2 \text{ are indep.}) \\
    & \quad \times P(\text{first } K-2 \text{ cols are indep.})
\end{align*}

Expanding this further, we get,
\begin{align*}
    P(\text{full rank}) &= \prod_{i=1}^{K} P(\text{col } i \text{ is indep.} | \text{ first } i-1 \text{ are indep.})
\end{align*}

Next, we will find $P(\text{col } i \text{ is indep.} | \text{ first } i-1 \text{ are indep.})$ and plug it back into the previous equation. Next,
\begin{align*}
    P(\text{col } i \text{ is indep.} | \text{ first } i-1 \text{ are indep.}) \\
    &= \frac{P^K - P^{i-1}}{P^K} \\
    &= 1 - \frac{P^{i-1}}{P^K}
\end{align*}

This is because $P^K$ is the total number of possibilities for the new column since each of the K entries can take any value from 0 to $P-1$. Out of these, we subtract the number of entries that will be a linear combination of the previously selected $i-1$ columns. Each such linear combination can be written as $\alpha_1 c_1 + \alpha_2 c_2 + \cdots + \alpha_{i-1} c_{i-1}$, where $c_1, c_2, \cdots, c_{i-1}$ are the $i-1$ columns and $\alpha_1, \alpha_2, \cdots, \alpha_{i-1}$ are the coefficients the columns are multiplied by to generate the linear combination. Each distinct value of the coefficients $\alpha_1, \alpha_2, \cdots, \alpha_{i-1}$ gives a distinct vector (because the previous columns are linearly independent), and there are $P^{i-1}$ such possibilities.

Substituting this back into the equation for P(full rank), we get,
\begin{align*}
P(\text{full rank}) &= \prod_{i=1}^{K} \left(1 - \frac{P^{i-1}}{P^K}\right) \\
&= \left(1 - \frac{P^0}{P^K}\right) \left(1 - \frac{P^1}{P^K}\right) \cdots \left(1 - \frac{P^{K-1}}{P^K}\right)
\end{align*}
Re-indexing with $j = K-(i-1)$, this product simplifies to:
\begin{equation*}
P(\text{full rank}) = \prod_{j=1}^{K} \left(1 - \frac{1}{P^j}\right)
\end{equation*}

Finally, the probability of being rank deficient is the complement:
\begin{equation*}
P(\text{rank deficient}) = 1 - P(\text{full rank}) = 1 - \prod_{i=1}^{K} \left( 1 - \frac{1}{P^{i}} \right)
\end{equation*}

\section{Intersection of Vector Spaces}
\label{app:subspace_intersection}
Additionally, calculating the intersection of two subspaces can be done by solving a system of linear equations. Therefore, it is computationally feasible as it can be done in polynomial time. We briefly discuss how one can compute the intersection of two sets of vectors $U = \{u_1, u_2, \cdots, u_m\}$ and $W = \{w_1, w_2, \cdots, w_n\}$ consisting of $n$ and $m$ vectors, respectively. Suppose a vector, $v$, is in the intersection of the subspaces spanned by $U$ and $W$. Then it can be written as a linear combination of the vectors in both $U$ and $W$. Thus, we can write,
\begin{equation}\label{eq:subspace_intersection}
v = \gamma_1 u_1 + \gamma_2 u_2 + \cdots + \gamma_m u_m = \delta_1 u_1 + \delta_2 u_2 + \cdots + \delta_n u_n
\end{equation}
where $\gamma_1, \gamma_2, \cdots, \gamma_m$ and $\delta_1, \delta_2, \cdots, \delta_n$ are the scalar coefficients of the respective linear combinations. To find all such $v$, one needs to find all the solutions to Equation~\ref{eq:subspace_intersection} and plug them back in to get $v$. Solving Equation \ref{eq:subspace_intersection} is equivalent to solving the system of linear equation below,
\begin{align*}
    \gamma_1 u_1 + \gamma_2 u_2 + \cdots + \gamma_m u_m = \delta_1 u_1 + \delta_2 u_2 + \cdots + \delta_n u_n \\
    \gamma_1 u_1 + \gamma_2 u_2 + \cdots + \gamma_m u_m - \delta_1 u_1 - \delta_2 u_2 - \cdots - \delta_n u_n = 0 \\
\end{align*}
This can be condensed into the matrix format as
\begin{align*}
    U \vec{\gamma} - W \vec{\delta} &= 0 \\
    \begin{bmatrix}
        U & W
    \end{bmatrix} \begin{bmatrix}
        \vec{\gamma} \\
        \vec{\delta}
    \end{bmatrix} &= 0 \\
\end{align*}
This can be solved by using standard linear algebra techniques like Gaussian Elimination. The solutions can then be plugged back into $v = U \vec{\gamma}$ to get all possible $v$, which will give the intersection subspace.

\begin{figure}[t]
  \centering
  \includegraphics[width=0.9\linewidth]{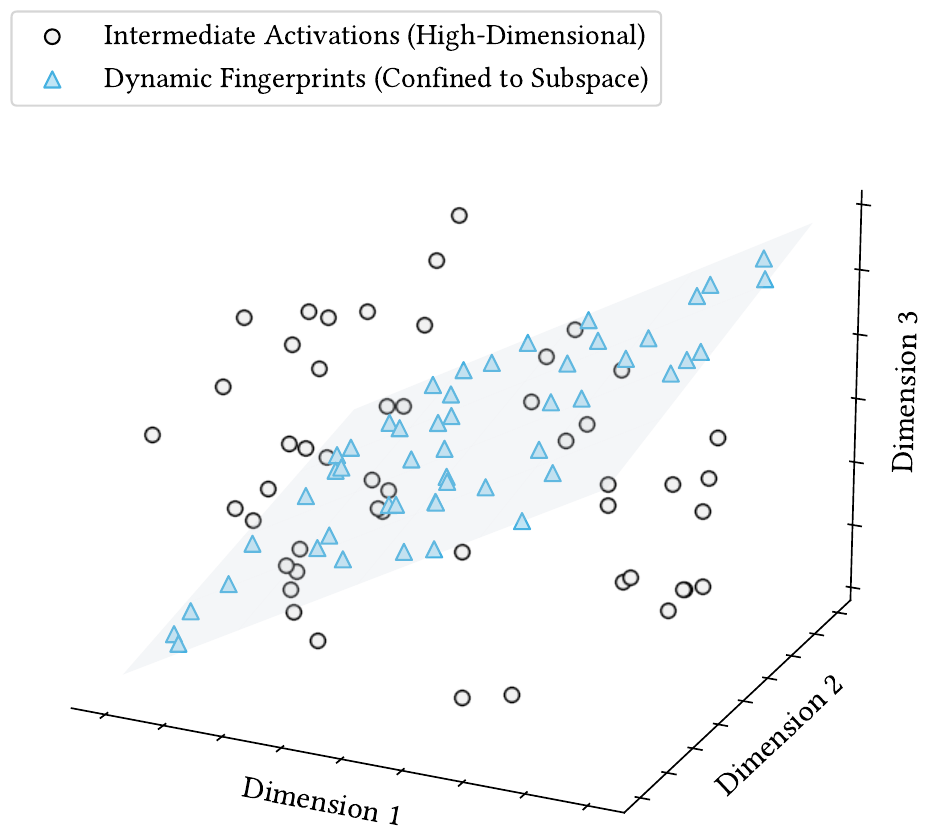}
  \caption{A visualization of the root cause of the vulnerability in Soter's integrity checking mechanism. The generated fingerprints (blue triangles) are confined to a low-dimensional subspace (gray plane) and are thus distinguishable from high-dimensional activations (gray circles).}
  \label{fig:subspace_vulnerability}
\end{figure}

\section{Discussion}
\label{sec:discussion}

The successful attacks detailed in Sections \ref{sec:direct_subspace_attack} and \ref{sec:cross_attack_soter} are more than specific flaws; they are practical demonstrations of a fundamental vulnerability pattern that arises when TEE-based systems prioritize performance via precomputation. The reuse of a static secret across multiple queries creates a structural dependency that allows an adversary to aggregate outputs and break the system. This section discusses the scope of our findings and outlines potential paths forward for designing more robust systems.

\subsection{Limitations and Scope}
It is important to clarify the boundaries of our work. First, our threat model assumes the TEE itself is a secure black box. Our attacks are at the protocol level and do not consider hardware side-channel vulnerabilities that could leak secrets directly from the enclave.

Second, we do not analyze adaptive or behavioral countermeasures. A system could, in theory, monitor for anomalous query patterns (such as repeated zero-vector inputs) and respond by rate-limiting or terminating a session. While such defenses could make our attacks more difficult to execute, they would not fix the underlying cryptographic flaw in the protocol.

\subsection{Paths Forward and Future Directions}
\label{sec:discussion_pathsforward_future}

Our findings motivate the search for secure, high-performance alternatives to the static basis pattern. One potential mitigation is \textbf{remote noise generation}, where a trusted server provides the on-device TEE with batches of single-use, precomputed noise effects. While this approach solves the static basis vulnerability, it has significant drawbacks: the system is no longer self-contained, and it becomes dependent on network availability and performance.

This analysis, combined with the failures of the systems we studied, leads to a set of core principles for future designs and highlights a key challenge for the community.

\subsubsection{Design Principles}
\paragraph{1. Eliminate Reused Static Secrets.} The central lesson of our work is that any secret material reused across queries is a potential liability. Future protocols must prioritize cryptographic designs that ensure forward secrecy by using a dynamic and non-reconstructible basis for each query.

\paragraph{2. Ensure Obfuscation Resists Cross-Query Analysis.} The attack on Soter proves that simply mixing a static secret basis with other high-dimensional data is an inadequate defense. A robust defense must ensure that no learnable structure persists across interactions that an adversary can aggregate.

The vulnerability of protocols that rely on a static secret basis motivates the search for a secure alternative. Any robust solution must use a sufficiently dynamic, non-reconstructible noise basis $\mathbf{N}_i$ for each query. The core difficulty lies in how the TEE can obtain the corresponding noise effect $\mathbf{N}_iW'$ efficiently. We discuss two primary mitigation paths for achieving this, each with significant trade-offs.

\subsubsection{Open Research Challenge}
The core difficulty in this domain remains reconciling the tension between performance and security. The ultimate goal is to design a protocol that is simultaneously performant and uses fresh randomness for every query. Achieving this will likely require new cryptographic primitives or hardware-software co-designs that can efficiently compute the effect of fresh noise without the prohibitive cost of on-the-fly matrix multiplication. Solving this challenge remains a critical direction for future research.

\section{Algorithms}
\label{app:algorithms}

\hrule 
\vspace{2pt}
{ 
    \captionsetup{skip=1pt} 
    \captionof{algorithm}{Direct Subspace Characterization Attack}
    \label{alg:perm_recovery_tlg_for_paper}
} 
\hrule
\medskip
\begin{algorithmic}[1] 

\Statex \textbf{Input:}
\Statex \hspace{\algorithmicindent} $\text{TEE}(\mathbf{a})$ that returns $\mathbf{y} \in \mathbb{F}_P^{d}$ for input $\mathbf{a} \in \mathbb{F}_P^{d}$.
\Statex \hspace{\algorithmicindent} $K$: Number of precomputed noise vectors.
\Statex \hspace{\algorithmicindent} $d$: Dimensionality of embedding vectors.
\Statex \hspace{\algorithmicindent} $P$: A prime such that integers mod P form $\mathbb{F}_P$.
\Statex \textbf{Output:}
\Statex \hspace{\algorithmicindent} Recovered permutation $\hat{\pi}: \{0, \dots, dim-1\} \to \{0, \dots, dim-1\}$.
\Statex

\State \textit{/* Generate the subspace $S_{obs}$ spanned by the K permuted noise vectors */}
\State $Y_{samples} \leftarrow \emptyset$
\State $\mathbf{a}_{zero} \leftarrow \mathbf{0} \in F_P^{dim}$
\For{$q \leftarrow 1$ \textbf{to} $K$}
    \State \textit{/* Sample a random vector from subspace $S_{obs}$. */}
    \State \textit{/* $\text{TEE}(\mathbf{a}_{zero}) = \pi (\mathbf{a}_{zero} + \mathbf{m}) = \pi \, \mathbf{m}$\,, where $\mathbf{m}$ is a */}
    \State \textit{/* random linear combination of the noise vectors. */}
    \State $\mathbf{y}_q \leftarrow \text{TEE}(\mathbf{a}_{zero})$
    \State Add $\mathbf{y}_q$ to $Y_{samples}$
\EndFor
\State $\mathcal{B}_{S_{obs}} \leftarrow \Call{ComputeBasis}{Y_{samples}}$
\Statex
\State \textit{/* Compute a projection to the orthogonal complement of the subspace $S_{obs}$ */}
\State $\text{Proj}_{S_{obs}^{\perp}} \leftarrow \Call{CreateOrthogonalComplement}{\mathcal{B}_{S_{obs}}}$
\State Initialize $\hat{\pi}[0 \dots dim-1] \leftarrow -1$
\For{$j \leftarrow 0$ \textbf{to} $dim-1$} 
    \State $\mathbf{e}_j \leftarrow \Call{CreateCanonicalVector}{j, d}$
    \State $\mathbf{y}_j \leftarrow \text{TEE}(\mathbf{e}_j)$
    \State $\mathbf{s}_j \leftarrow \text{Proj}_{S_{obs}^{\perp}}(\mathbf{y}_j)$
    \State \textit{/* Projection cancels noise: */}
    \State \textit{/* $\mathbf{s}_j = \text{Proj}_{S_{obs}^{\perp}}(\mathbf{y}_j) = \text{Proj}_{S_{obs}^{\perp}}(\pi(\mathbf{e}_j + \mathbf{m})) $ */}
    \State \textit{/* \quad $= \text{Proj}_{S_{obs}^{\perp}}(\pi \, \mathbf{e}_j) \qquad \because \pi \, \mathbf{m} \in S_{obs} $*/}
    \State $\hat{\pi}[j] \leftarrow \Call{FindColumnIndex}{\text{Proj}_{S_{obs}^{\perp}}, \mathbf{s}_j} $
\EndFor
\Statex
\State \Return $\hat{\pi}$
\end{algorithmic}
\vspace{2pt}
\hrule 
\bigskip 

\hrule 
\vspace{2pt}
{ 
    \captionsetup{skip=1pt} 
    \captionof{algorithm}{Integrity Bypass Attack on Soter-like Systems}
    \label{alg:soter_attack}
} 
\hrule
\medskip
\begin{algorithmic}[1] 

\Statex \textbf{Input:}
\Statex \hspace{\algorithmicindent} $\text{ObserveInferenceBatch}()$ that returns a batch of vectors from the TEE.
\Statex \hspace{\algorithmicindent} $k$: The dimension of the secret cornerstone basis.
\Statex \hspace{\algorithmicindent} $d$: The dimensionality of the vectors.

\Statex
\Statex \textit{/* Stage 1: Recover Cornerstone Fingerprint Subspace $V_C$ */}
\State $\text{CollectedVectorsA} \leftarrow \emptyset$
\State $\text{CollectedVectorsB} \leftarrow \emptyset$
\State \textit{/* Passively observe $k+\delta$ batches for each set to ensure they span the subspace */}
\For{$i \leftarrow 1$ \textbf{to} $k+\delta$}
    \State $\text{batch}_A \leftarrow \Call{ObserveInferenceBatch}{}$
    \State $\text{batch}_B \leftarrow \Call{ObserveInferenceBatch}{}$ 
    \State Add all vectors from $\text{batch}_A$ to $\text{CollectedVectorsA}$
    \State Add all vectors from $\text{batch}_B$ to $\text{CollectedVectorsB}$
\EndFor
\State $\mathcal{B}_{U_A} \leftarrow \Call{ComputeBasis}{\text{CollectedVectorsA}}$
\State $\mathcal{B}_{U_B} \leftarrow \Call{ComputeBasis}{\text{CollectedVectorsB}}$
\Statex
\State \textit{/* The cornerstone subspace is the intersection of the vector spaces spanned by the two sets. */}
\State $\mathcal{B}_{V_C} \leftarrow \Call{SubspaceIntersection}{\mathcal{B}_{U_A}, \mathcal{B}_{U_B}}$
\Statex

\State \textit{/* Stage 2: Identify Fingerprints to Bypass Integrity Checks */}
\Statex
\Function{BypassIntegrityCheck}{$\text{new\_batch}$, $\mathcal{B}_{V_C}$}
    \For{each $\mathbf{v}$ \textbf{in} $\text{new\_batch}$}
        \If{\Call{IsInSubspace}{$\mathbf{v}$, $\mathcal{B}_{V_C}$}}
            \State \textit{/* Vector is a fingerprint; process it correctly to pass the check. */}
            \State $\Call{ReturnCorrectResult}{\mathbf{v}}$
        \Else
            \State \textit{/* Vector is genuine data; return a tampered result. */}
            \State $\Call{ReturnTamperedResult}{\mathbf{v}}$
        \EndIf
    \EndFor
\EndFunction

\end{algorithmic}
\vspace{2pt}
\hrule 

\end{document}